%% file: main_ESTES_overview.tex
\documentclass[%
 reprint,
 amsmath,
 amssymb,
 aps,
 prd,
 superscriptaddress,
 nolongbibliography
 ]{revtex4-2}

\usepackage{graphicx}
\usepackage{float}
\usepackage{dcolumn}
\usepackage{bm}
\usepackage[caption=false]{subfig}
\usepackage{natbib}
\usepackage{multirow}
\usepackage{xcolor}
\usepackage{hyperref}

\begin{document}

\preprint{PRD/1234}

\title{Characterization of the Astrophysical Diffuse Neutrino Flux using Starting Track Events in IceCube}

\date{\today}

\begin{abstract}
A measurement of the diffuse astrophysical neutrino spectrum is presented using IceCube data collected from 2011-2022 (10.3 years). We developed novel detection techniques to search for events with a contained vertex and exiting track induced by muon neutrinos undergoing a charged-current interaction. Searching for these starting track events allows us to not only more effectively reject atmospheric muons but also atmospheric neutrino backgrounds in the southern sky, opening a new window to the sub-100 TeV astrophysical neutrino sky. The event selection is constructed using a dynamic starting track veto and machine learning algorithms. We use this data to measure the astrophysical diffuse flux as a single power law flux (SPL) with a best-fit spectral index of $\gamma = 2.58 ^{+0.10}_{-0.09}$ and per-flavor normalization of $\phi^{\mathrm{Astro}}_{\mathrm{per-flavor}} = 1.68 ^{+0.19}_{-0.22} \times 10^{-18} \times \mathrm{GeV}^{-1} \mathrm{cm}^{-2} \mathrm{s}^{-1}  \mathrm{sr}^{-1}$ (at 100 TeV). The sensitive energy range for this dataset is 3 - 550 TeV under the SPL assumption. This data was also used to measure the flux under a broken power law, however we did not find any evidence of a low energy cutoff. 
\end{abstract}

\include{authorlist}
\collaboration{IceCube Collaboration}
\noaffiliation

\maketitle

\tableofcontents

\section{\label{sec:Introduction} Introduction}
High energy astrophysical neutrinos were discovered by IceCube in 2013 \cite{IceCube:2013low,PhysRevLett.111.021103, IceCube:2014stg}; 
since then, there have been many efforts to understand the production mechanisms by which these high energy neutrinos are created. 
Energetic neutrinos are decay products, and the neutrino flux points directly to the processes that created it \cite{Workman:2022ynf}. 
In particular, we know there are extremely energetic accelerators which drive cosmic rays to very high energies via processes such as Fermi acceleration \cite{PhysRev.75.1169,1977DoSSR.234.1306K,10.1093/mnras/182.2.147}. 


It is expected that some cosmic rays will interact with hadronic matter or the photon flux  near their source.  Charged pions $(\pi^{\pm})$ and kaons $(\mathrm{K}^{\pm})$, produced in these interactions decay into neutrinos and muons ($\pi^{\pm} \rightarrow \nu_\mu (\bar\nu_\mu) + \mu^{\pm}$), with the muons subsequently decaying into electrons and neutrinos ($\mu^{\pm} \rightarrow \mathrm{e}^{\pm} + \bar\nu_\mathrm{e} (\nu_\mathrm{e}) + \nu_\mu (\bar\nu_\mu)$). In this scenario, the neutrino flavor ratio at the source is $\nu_\mu:\nu_\mathrm{e}:\nu_\tau=2:1:0$ with oscillations converting this ratio to approximately 1:1:1 \cite{Becker:2007sv} at Earth, although there are alternative scenarios which predict other ratios \cite{PhysRevLett.95.181101}.

\begin{figure*}
\includegraphics[width=0.32\textwidth]{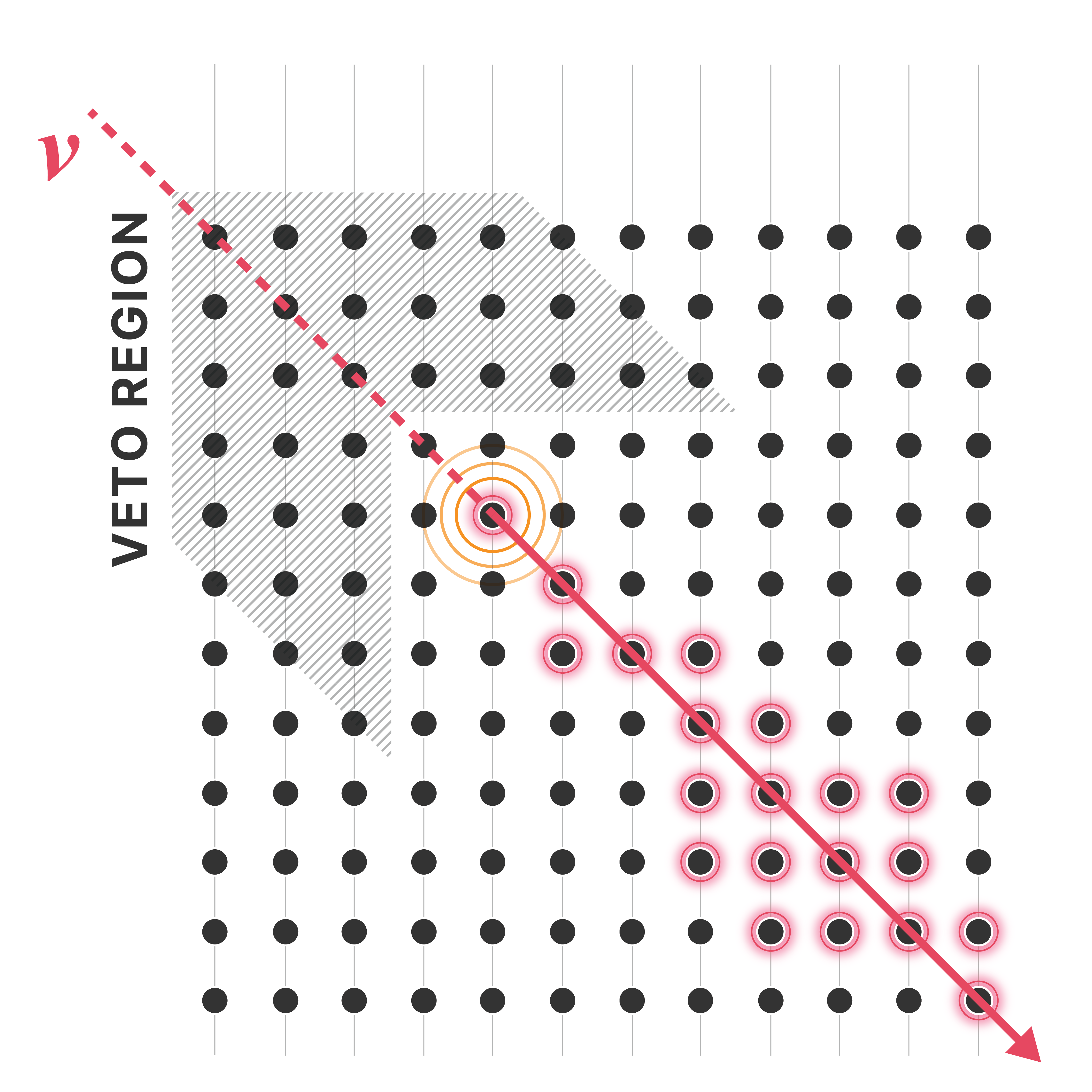}
\includegraphics[width=0.32\textwidth]{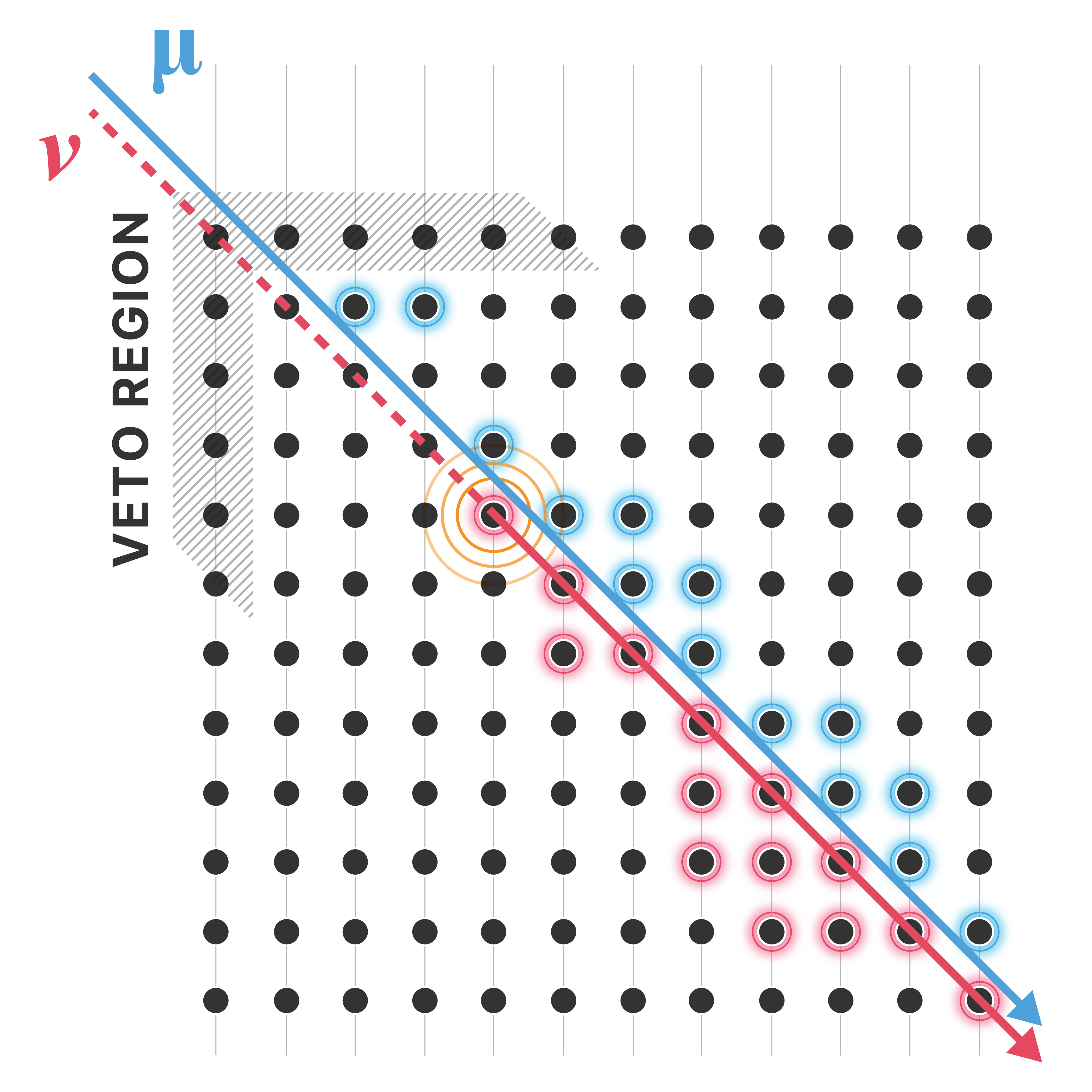}
\includegraphics[width=0.32\textwidth]{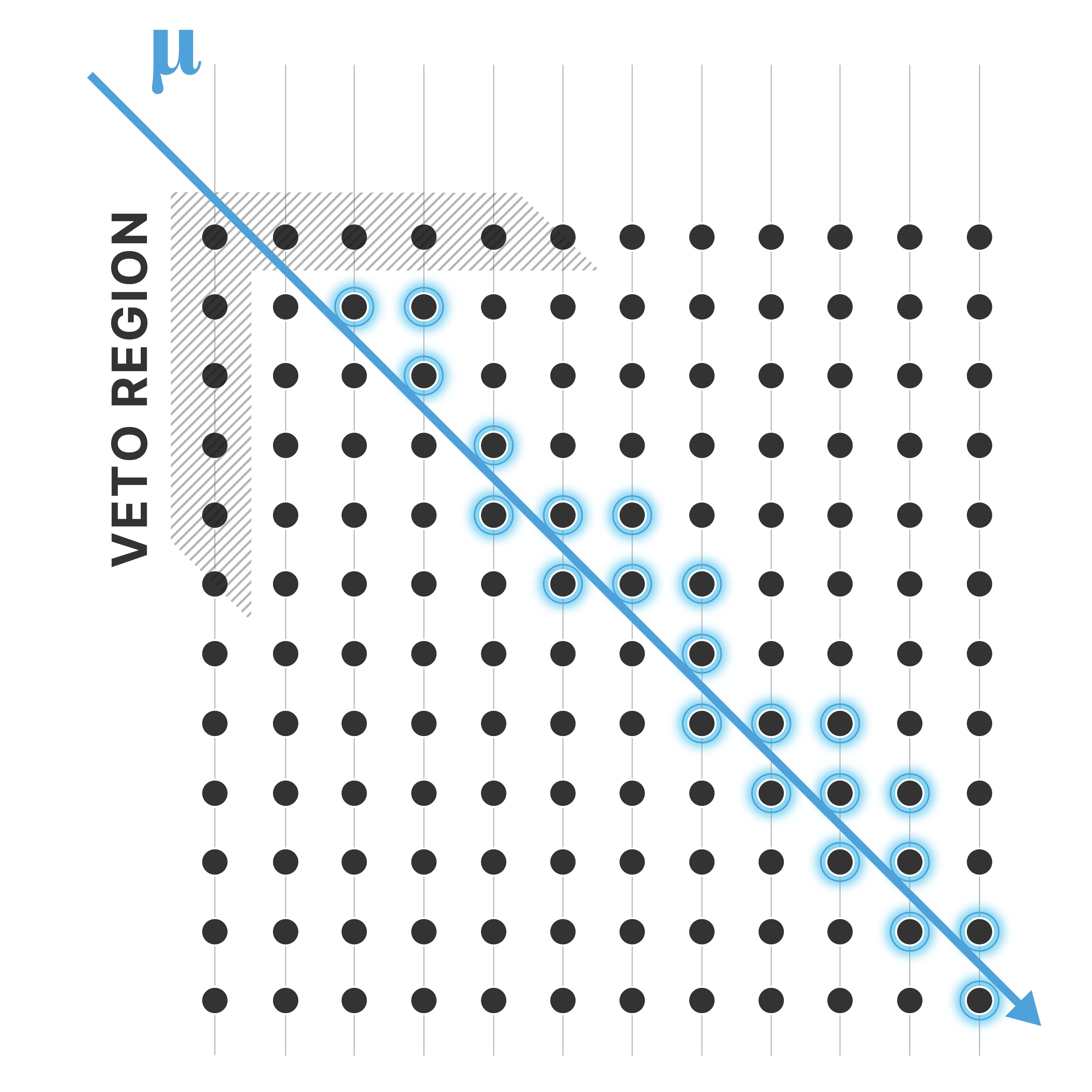}
\caption{The diagram illustrates the concept of a dynamic veto region described in Sec. \ref{sec:STV}. The depth of the veto region depends on the location of the earliest hits and the total number of hits for a particular event. The orange circles indicate the presence of the hadronic shower from the muon neutrino interaction. 
Left: A muon neutrino interacts inside the detector. It leaves no light inside the veto region; therefore, the event is accepted. Center and right: An atmospheric muon will deposit light near the detector edge resulting in a much smaller veto region and high likelihood of light observed within the veto region (blue dots). This is true for events with a single muon (right), muon bundles, and atmospheric neutrinos with a coincident muon from the same air shower (center). These are rejected by the ESTES data selection.}
\label{fig:startingtrack}
\end{figure*}

In this paper, we present a measurement of the diffuse astrophysical neutrino flux using novel classification techniques and methods to reconstruct event observables. The event selection and techniques described in this paper are referred to as the Enhanced Starting Track Event Selection (ESTES) \cite{KyleThesis,Silva:2019fnq,Mancina:2019hsp,Silva_2021,Mancina_2021,IceCube:2021ctg,Silva:2023wol,SarahThesis,ManuelThesis}. The selection criteria reject atmospheric muons and atmospheric neutrinos with accompanying muons in the southern equatorial sky, extending the measurement of the astrophysical diffuse flux down to 3 TeV. 

Section \ref{sec:Overview} provides an overview of this paper's purposes and goals. Section \ref{sec:detsim} discusses the detector configuration and the simulated data used in this measurement. Section \ref{sec:ReconstructedObservables} outlines how the neutrino energy and direction are reconstructed. Section \ref{sec:EventSelection} is a summary of the event selection. Section \ref{sec:measurement} summarizes the likelihood techniques employed and how systematic uncertainties are incorporated. This is a binned-likelihood analysis based on expectations from simulated astrophysical neutrinos, atmospheric neutrinos and muons. Finally, Section \ref{sec:Results} discusses the results of the single power law, broken power law, hemisphere model, and unfolded flux measurements. A companion search for neutrino sources using the ESTES data selection is presented in an accompanying paper \cite{ESTESNS}.

\section{\label{sec:Overview} Measurement Motivation}
\subsection{\label{sec:physics} Astrophysical neutrinos}
IceCube searches for neutrino sources have seen evidence for neutrino emission from TXS 0506+056 \cite{IceCube:2018cha,IceCube:2018dnn}, NGC 1068 \cite{NGC1068}, and the Milky Way \cite{dnncascade}. However, the flux measured from these three sources is only a small part of the total observed diffuse flux. Additional neutrino sources, potentially from multiple populations, are required to explain it in full \cite{Murase:2016gly, PhysRevD.101.123017, IceCube:2022ham}.

IceCube finds the total astrophysical neutrino diffuse flux to be generally well described by a single power law, and no additional complexity has so far been established. However, there are reasons to believe that cosmic-ray accelerators could produce spectral features at TeV and sub-TeV energies, which motivates further detailed study of the diffuse flux \cite{Fang:2022trf}. 

In pp-scenarios, 
the cosmic rays interact with gas near the acceleration site. 
Neutrinos produced from pions and kaons 
follow the energies of cosmic rays, and the neutrino flux is expected with a similar spectral index as these cosmic rays \cite{Murase:2015xka}. A hardening of the flux is predicted below a break energy in specific models (motivated by cosmic ray diffusion), the neutrino flux is only expected to harden to a spectral index ($\gamma$) of $\sim2.0$ below this break \cite{Loeb:2006tw,Murase:2008yt}. 

In  p$\gamma$ scenarios, the cosmic rays interact with a photon gas near the production site. This has been suggested to occur in cosmic ray reservoirs such as active galaxies (e.g. in NGC 1068) and other types of p$\gamma$ sources \cite{Berezinsky:867657,1979ApJ...232..106E,1986ApJ...304..178K,1990ApJ...362...38B,1991ApJ...380L..51H,PhysRevLett.69.2738,Szabo:1994qx,Alvarez-Muniz:2004xlu,Cuoco:2007aa,Koers:2008hv,Jacobsen:2015mga,Murase:2015ndr}. The properties of the photon gas, in particular, the optical depth to photo-meson production, therefore drive the properties of the expected neutrino flux. 

In either case, pp or p$\gamma$, the charged pions/kaons and muons can be further accelerated resulting in a hardening in the neutrino spectrum \cite{PhysRevLett.93.181101,PhysRevLett.95.061103,Koers:2007je,Klein:2012ug,PhysRevD.88.121301}. 
 
\subsection{\label{sec:estesnew} Starting track morphology}

While astrophysical neutrinos are the target of this analysis, the largest contributors to the ESTES dataset below 100\,TeV are atmospheric muons and atmospheric neutrinos. Atmospheric muons trigger the detector at a rate of 3000 Hz \cite{AARTSEN20161}. In comparison, approximately 100 astrophysical neutrinos are expected per year in this dataset. To improve signal purity, a series of complex cuts is deployed as described in detail in Section \ref{sec:EventSelection}.

Examples of strategies used recently by IceCube to measure the astrophysical diffuse flux are: a cascade dominated measurement \cite{SBUCasc}, the selection of muon neutrinos from the northern sky \cite{IceCube:2021uhz}, and the ``starting event” selections \cite{HESENew, IceCube:2014rwe, IceCube:2014rwe,inelasticity}. The northern sky tracks dataset applies a cut in zenith to reject the overwhelming background from atmospheric muons. However, this data set is still dominated by atmospheric neutrinos at energies below 100\,TeV.  

One way to distinguish incoming neutrinos from downgoing muons uses an event signature where the interaction vertex can be located inside the detector. In contrast, incoming muons are removed if they have early photons recorded in the outer regions of the detector. 
The starting events selections \cite{HESENew, IceCube:2014rwe, IceCube:2014rwe,inelasticity} reduce the muon rate in the southern sky through veto techniques, whereby events are removed if they have early photons recorded in the outer regions of the detector. Retained events in \cite{HESENew, IceCube:2014rwe, IceCube:2014rwe,inelasticity} include both cascade and starting track events.
These veto-based datasets also take advantage of the neutrino self-veto effect \cite{PhysRevD.79.043009,PhysRevD.90.023009,NuVeto}, which results in a suppression 
not only of atmospheric muons, but also of the atmospheric neutrinos in the southern sky due to the removal of atmospheric neutrinos that are accompanied by 
muons from the same shower.  
The self-veto was first implemented by the High Energy Starting Events (HESE, \cite{HESENew}) analysis. A more complex veto was later constructed for the Medium Energy Starting Events (MESE, \cite{IceCube:2014rwe}) analysis, which applies a veto volume proportional to the charge of the event (lower charge, greater veto region size). While these datasets also allow for starting tracks, the rates are greatly reduced in the southern sky at lower energies due to their strict veto definitions.  

The ESTES dataset takes advantage of the muon track topology to apply a dynamic veto and machine learning to greatly improve retention of starting track events in the southern sky.
The dynamic veto is computed event-to-event using the position and direction of the first observed photon. The machine learning algorithms then use the distribution of the energy losses along the muon track and the positions to estimate the probability of a particular event being an astrophysical neutrino. This method allows measuring the astrophysical muon neutrino flux at lower energies. Figure \ref{fig:startingtrack} illustrates the differences between the ESTES target event morphology (left) and background morphologies (middle and right) present in the analysis.

Starting track events occur when a muon neutrino undergoes a charged current deep inelastic scattering interaction within the fiducial, or interior, volume of the detector. An initial cascade is observed from the hadronic component of the interaction followed by a muon track that eventually exits the detector. The presence of the cascade is advantageous as it gives us more access to the neutrino energy because a higher proportion of the neutrino's energy is deposited inside the detector. The exiting muon track is also useful since it is then used to reconstruct the neutrino direction.

\begin{figure}[t!]
\includegraphics[width=0.49\textwidth]{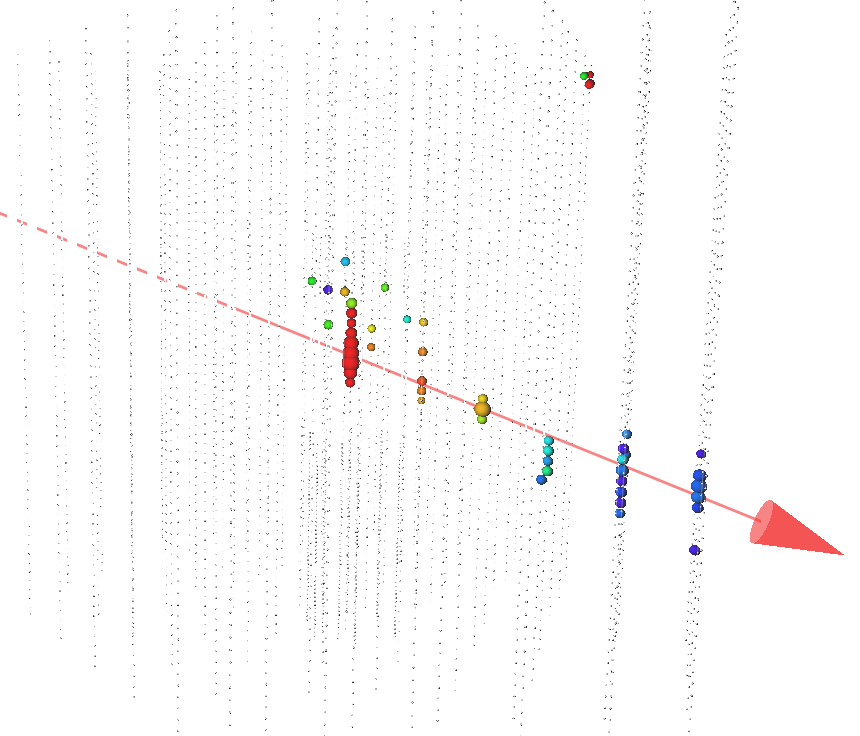}
\caption{Event display of a data event satisfying the criteria for the ESTES data selection. Red represents the earliest light detected, while blue represents the last light detected. The reconstructed zenith angle is $\theta = 71^{\circ}$, and the reconstructed neutrino energy (sum of the predicted cascade and muon energies) is E = 11 TeV. The dashed line is the reconstructed trajectory of the neutrino and the solid line is that of the reconstructed muon. The cluster of early hits (in red) is the cascade component, and the line of deposited charge is the track component. No charge is deposited in the outer layers (upper left) of the detector, which makes this a ``starting" event.}
\label{fig:sampleevent}
\end{figure}

An example of a starting track data event, which passed all cuts, is shown in Fig. \ref{fig:sampleevent}. This event's reconstructed zenith angle is 71$^{\circ}$ and reconstructed neutrino energy, defined as the sum of the predicted cascade and muon energies, is 11\,TeV respectively using the techniques discussed in Sec. \ref{sec:ReconstructedObservables}.

\section{\label{sec:detsim} Detector and Simulations}
\subsection{\label{sec:detector} Detector configuration}
The IceCube Neutrino Observatory is a cubic-kilometer sized detector located in the geographic South Pole buried 1.5\,km under the Antarctic ice \cite{holeice}. The detector is comprised of 5160 digital optical modules (DOMs), which each consist of a single photomultiplier tube (PMT) \cite{icecubepmt} and associated data acquisition electronics \cite{icecubedaq}. As relativistic charged particles traverse the ice, the particles emit Cherenkov photons \cite{cherenkov}. The DOMs will detect some of these photons, which are then converted by the readout system into an electronic signal. We refer to a discrete signal in units of photo-electrons (PEs) and assign it a position and time. 
 
The detector consists of a hexagonal grid of 86 instrumented cables referred to as strings \cite{icecubedaq}. The DOMs are spaced 17\,m apart vertically on the strings, and the strings have a horizontal separation of 125\,m. There is a central array, known as DeepCore, with 8 strings spaced about 70\,m apart with each string containing 60 high quantum efficiency DOMs spaced about 7\,m apart \cite{IceCube:2011ucd}. We use IceCube data collected from 2011-2022 and select runs where the entire 86-string detector was operational. 

\subsection{\label{sec:Flux} Simulation of neutrinos and atmospheric muons}
\begin{figure}[t!]
\includegraphics[width=0.45\textwidth]{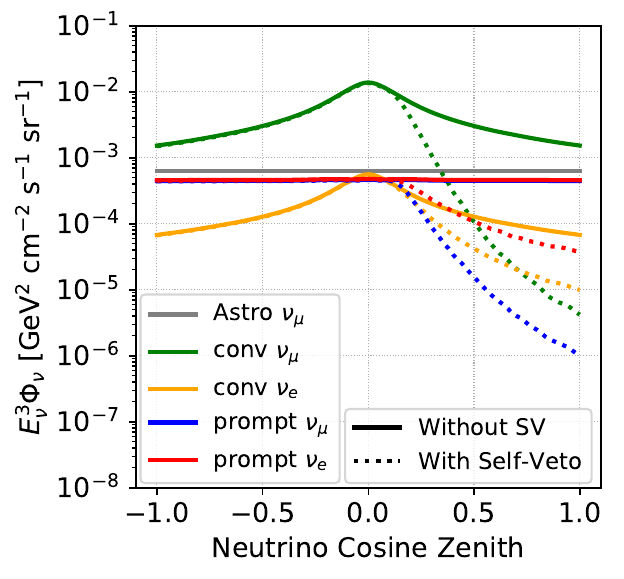}
\caption{Atmospheric neutrino fluxes with and without accounting for the self-veto. We observe a strong suppression of down-going atmospheric neutrinos. We note that the astrophysical neutrino rate is higher than the atmospheric neutrino rate for events that pass the selection with energies of 50 TeV and above for cosine zeniths greater than 0.2 with the self-veto applied. (Note: matter effects not included here)}
\label{fig:fluxes}
\end{figure}

To model the atmospheric muons and neutrinos, produced through the interaction of cosmic rays with the atmosphere, we use the Gaisser H4a cosmic-ray model \cite{H4a} and the Sibyll 2.3c hadronic interaction model \cite{23c} as a baseline. We use the Matrix Cascade Equation solver (MCEq) software package \cite{MCEq} to compute the fluxes at the Earth's surface. We define the conventional neutrino flux as the neutrinos from the decay of pions, kaons, and muons as the cosmic ray showers evolve in the atmosphere. The prompt neutrino flux is defined as the neutrino flux produced by the decay of charmed hadrons \cite{Workman:2022ynf}. These particles decay promptly $(\tau \sim 10^{-12} \mathrm{s})$ in the atmosphere. The conventional and prompt neutrinos are shown as a function of zenith at 50 TeV in Fig.\ref{fig:fluxes}.  The treatment of systematic uncertainties in the modeling of these atmospheric backgrounds is discussed in Sec \ref{sec:theorysyst}. 

Interactions of neutrinos in the Earth are modeled assuming the Cooper-Sarkar-Mertsch-Sarkar (CSMS) neutrino-nucleon cross-section \cite{CSMS} and Preliminary Reference Earth Model (PREM) \cite{PREM} Earth density model. In addition, deep inelastic scattering interactions near (and inside) the detector are simulated using the NuGen software package \cite{NuGen} which calculates the probability of this interaction occurring. NuGen also computes the daughter particle properties such as energy and direction. 

For neutrinos with zenith angle $< 90^{\circ}$, down-going neutrinos from the southern sky, we take into account the ``self-veto" effect \cite{PhysRevD.79.043009,PhysRevD.90.023009,NuVeto} using the Nu-Veto software package \cite{NuVeto}. The self-veto effect is an analytical adjustment to the atmospheric neutrino flux after taking into account atmospheric muons from the same air shower and how efficiently we can tag and remove these types of events. We model the muon rejection probability using a Heaviside step function where all events containing a muon above a particular energy are rejected with a 100$\%$ probability. In this analysis the probability is modeled with an energy-dependent nuisance parameter as described in Sec. \ref{sec:theorysyst}. The self-veto effect is illustrated in Fig. \ref{fig:fluxes} using the SPL best fit $\eta_{\mathrm{Self-Veto}} = 2.1$ (126 GeV), as defined in Sec. \ref{sec:SystematicUncertainties}. 

The atmospheric muon events are simulated using MuonGun \cite{MuonGun} for single muons and CORSIKA \cite{CORSIKA} for muon bundles. MuonGun has the advantage of simulating targeted Monte Carlo (MC) where the muons are all simulated near the detector allowing us to generate a sufficiently large MC sample. CORSIKA simulates full cosmic ray air-showers. This is advantageous because we can model the multi-muon detector response. CORSIKA was used to model the muon rates in a background-dominated region to validate the event selection performance, further described in reference \cite{IceCube:2021ctg}, and MuonGun was used to model the remaining muon background after all cuts were applied. To prevent double counting of single muons between CORSIKA and MuonGun, we match all muons from a CORSIKA shower where the muons intersect with the MuonGun simulated detector geometry. If the CORSIKA shower consists of only a single muon, the event is removed.

The charged leptons are propagated through the South Pole ice using PROPOSAL \cite{PROPOSAL}. PROPOSAL models the energy losses of the muons and taus as they travel through the ice over extended distances. It also models stochastic processes such as inelastic photonuclear interactions where the secondary particles are also propagated through the ice. Cascade shower development is modeled using the Cascade Monte Carlo (CMC) program \cite{CMC}. CMC models the longitudinal development of the cascade-shower. The relativistic charged particles traverse the ice and emit photons. The photons are propagated through the ice assuming a South Pole ice depth-dependent scattering and absorption coefficient as described in Refs. \cite{SPICE1,SPICE2}. These ice models are then used to compute the expected arrival time and direction at any given DOM in the detector.

\section{\label{sec:ReconstructedObservables} Reconstructed Observables}

The morphology of the observed light is used to reconstruct the event energy and direction of the charged particles. The observables described in this section rely on previously published IceCube algorithms. Some modifications to the algorithms have been made, to take advantage of the mixed properties of starting track events which can include light from a hadronic cascade and muon track. We discuss these modifications and the expected performance. We also rely on these reconstruction algorithms as seeds to more complex observables used in the event selection. In Sec. \ref{sec:EventSelection}, these will explicitly be defined and labeled.

\subsection{\label{sec:direction}  Directional reconstruction}
The direction of the event is reconstructed using a series of increasingly complex algorithms. Initial directional reconstructions are performed using the LineFit \cite{linefit, AARTSEN2014143}, SPEFit \cite{splinempe}, SplineMPE \cite{splinempe}, and Millipede \cite{millipede} algorithms with each subsequent algorithm seeded by the track direction found using the previous algorithm. The most complex directional algorithm Millipede takes into account all active DOMs in the detector, regardless of whether they observe a signal or not, to compute the overall direction and energy of the track. Millipede works by splitting up the track into segments and fitting an independent energy loss for each segment, thus accommodating the stochastic nature of the energy loss of muons. Millipede is run once to calculate the track direction and a set of energy losses along the track. This is then used to compute many of the Starting Tracks BDT inputs as described in Sec. \ref{sec:BDT}. 

However, the hadronic cascade at the neutrino interaction vertex can introduce an incorrect initial track direction for events with shorter track lengths. Also, neutrino events with coincidental muons (not from the same air shower) often result in incorrect directions due to the presence of the photons from the second track. We, therefore, introduce a refined reconstruction that is a sequence of three algorithms: 1) iterative-SplineMPE \cite{splinempe} 2) f$_{\mathrm{hit}}$ algorithm to select the ``best track" from a list of tracks 3) Millipede using the ``best track" as a seed. 

The iterative-SplineMPE algorithm is adopted from reference \cite{splinempe}. This algorithm attempts to find a global zenith angle by refitting the track using different zenith angles, azimuth angles, and positions randomly selected and connecting it to the previous best-fit vertex. For each tested track, if the likelihood shows improvement, then this new track is chosen as the best track. If no improvement is seen over 50 iterations, then this iterative-SplineMPE is passed to the next step. 

\begin{figure}[t!]
\includegraphics[width=0.45\textwidth]{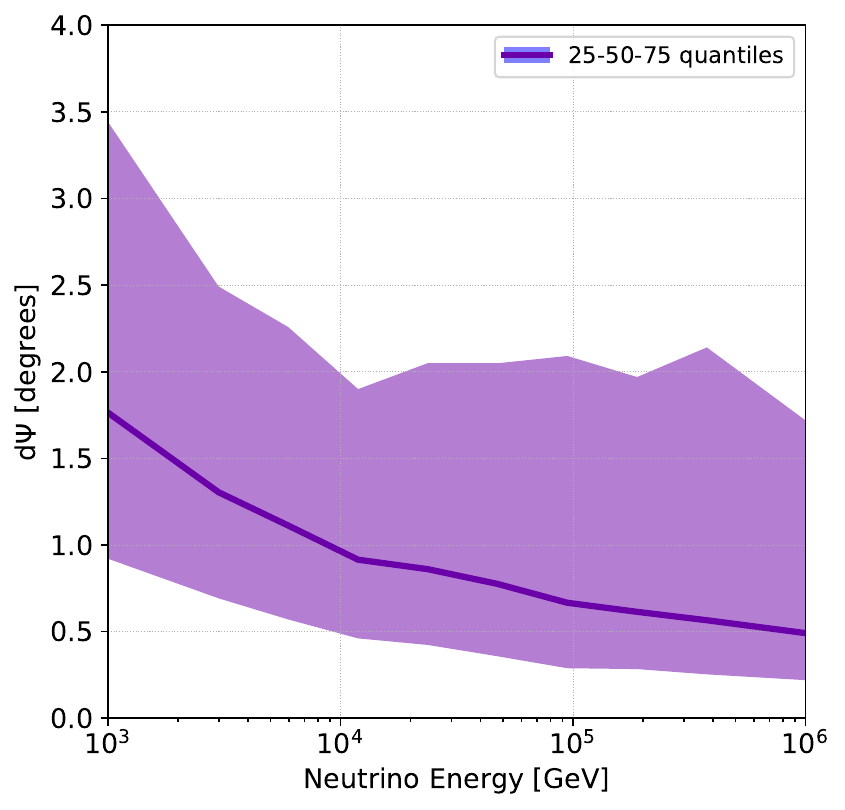}
\caption{Directional error 25, 50, 75$\%$ quantiles for starting muon neutrinos as a function of the real neutrino energy. d$\psi$ is the space angle distance between the reconstructed direction and the truth direction. At 1\,TeV the events are reconstructed with a median resolution of $1.6^{\circ}$, at 100\,TeV we see a median resolution of $0.66^{\circ}$.}
\label{fig:directionalreco}
\end{figure}

The quantity f$_{\mathrm{hit}}$ is then computed for a pre-defined list of tracks (LineFit, SPEFit, SplineMPE, iterative-SplineMPE, and Millipede). To calculate f$_{\mathrm{hit}}$, we take a reconstructed track hypothesis and find all DOMs within a perpendicular distance $r$ along the reconstructed track. This is a cylinder centered around the track with radius $r$. The f$_{\mathrm{hit}}$ is defined as the fraction of DOMs that detect at least one hit within this radius $r$ cylinder. The f$_{\mathrm{hit}}$ distribution is defined for different choices of $r$ (100m, 200m, etc...). The track with the greatest cumulative f$_{\mathrm{hit}}$ is selected as the ``best track". This ``best track" is then used as the seed to the Millipede algorithm again. The resulting direction from this Millipede fit is then used as the observable for the flux measurement. The angular resolution achieved using this procedure is shown in Fig. \ref{fig:directionalreco}. The directional resolution for this procedure is $1.5^\circ$ at 1 TeV and $0.66^\circ$ at 100 TeV for starting track events. 

\subsection{\label{sec:energy} Energy reconstruction}

\begin{figure}[t!]
\includegraphics[width=0.45\textwidth]{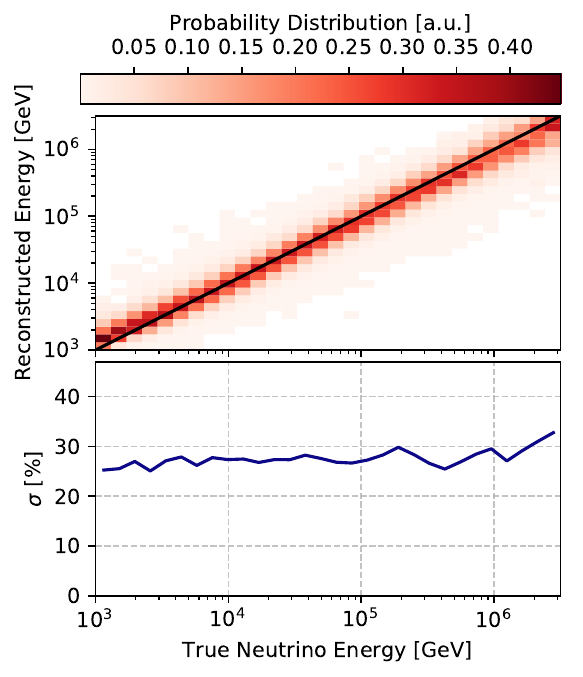}\\
\includegraphics[width=0.45\textwidth]{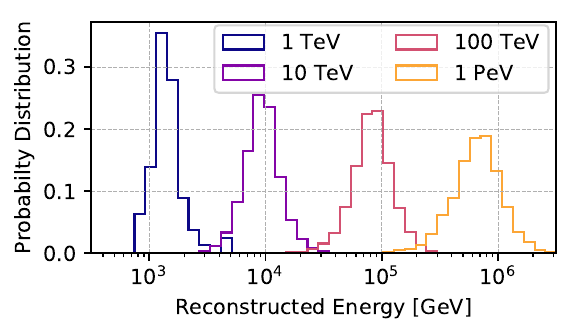}
\caption{Top: Reconstructed event energy shown as a function of true neutrino energy. For each true neutrino energy slice, the distribution of events is normalized. The energy resolution ($\sigma$) for starting track events is observed to be near 25\% for 1 TeV and worsens to 30\% above 1 PeV. This is due to the increased amount of energy the muon eventually exits the detector with for higher energy events. Bottom: Slices of reconstructed event energies for neutrinos with true energy 1 TeV, 10 TeV, 100 TeV, and 1 PeV. We observe a slight loss in energy resolution towards higher energies.}
\label{fig:energyresolution}
\end{figure}

The distribution of energy losses for the event is first calculated using the Millipede algorithm \cite{millipede} from Sec. \ref{sec:direction}. Millipede was set to compute the deposited energy every 10 m along the track direction. The energy loss per segment is then used to train a Random Forest \cite{Breiman:2001hzm, 10.1007/s10994-006-6226-1} to predict the energy from the hadronic and muonic components separately \cite{inelasticity}. Here, the muon energy is defined as the muon's energy at its creation. We take the sum of the predicted cascade and muon energies to define the ``reconstructed energy" of each event. Figure \ref{fig:energyresolution} shows the reconstructed energy as a function of the true neutrino energy assuming muon neutrino events only.

The energy resolution for this dataset is 25\% at 1\,TeV and remains almost constant up to $\sim$ 1\,PeV. The top panel of Fig. \ref{fig:energyresolution} shows a slight biasing towards reconstructing higher energies at 1\,TeV. Above 1 PeV, the energy resolution degrades to 30\%. This loss in resolution is due to the increasing amount of energy the muon escapes the detector with. 
However, the achieved energy resolution of $\sim25-30\%$ is 
a significant improvement over that which is traditionally achieved using track events \cite{2013NIMPA.703..190A,IceCube:2019cia,IceCube:2021uhz}. 

\section{\label{sec:EventSelection} Event Selection}
\subsection{\label{sec:precut} Quality cuts}
\begin{figure}[t!]
\includegraphics[width=0.45\textwidth]{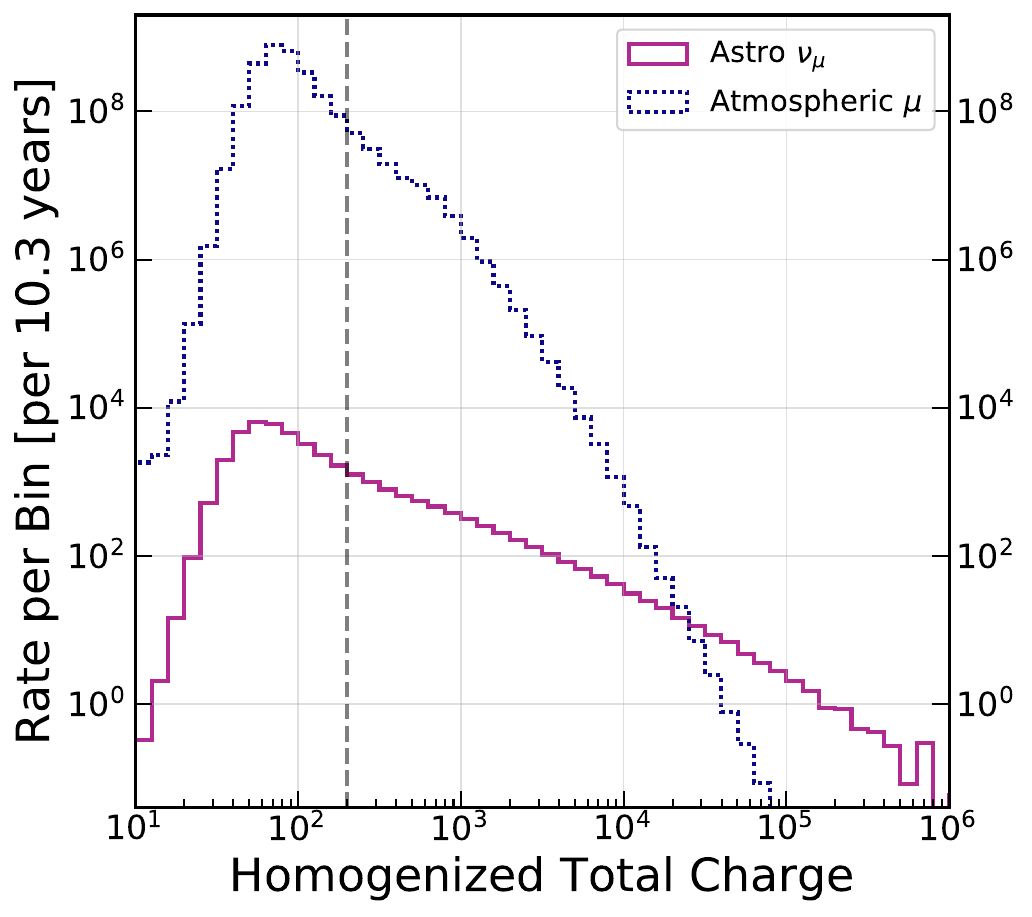}
\caption{Homogenized total charge (HQtot/PE) distribution for simulated atmospheric muons and astrophysical neutrinos.}
\label{fig:hqtotl2}
\end{figure}

Observed photons are recorded in units of PEs after taking into account quantum efficiency and calibration effects \cite{IceCube:2020nwx}. The IceCube detector trigger requires at least 8 locally coincident DOM hits recorded within a 5 microsecond window. After the trigger, an event is processed through the IceCube filtering scheme. There are many different filters which all involve fast selection criteria relying on basic event properties. Each filter targets different event morphologies. The ESTES dataset selects events which pass at least one of the following filters: Muon-Filter or Full-Sky-Starting-Filter \cite{ICRC2007}. One further quality cut is applied to the data set before the specific starting track event veto criteria are implemented. This cut is based on a quantity called the ``homogenized total charge" (HQTot). HQTot is calculated by summing all detected PEs, without DeepCore and excluding DOMs where the sum of PEs in a single DOM is more than 50\% of the total charge of the event because of their negative impact on reconstruction quality. The HQTot is shown in Fig. \ref{fig:hqtotl2} for simulated atmospheric muons and astrophysical neutrinos. The astrophysical neutrino rates are further split into ``contained" or ``uncontained vertex", which are defined as events with a simulated vertex located within or outside of the fiducial volume of the detector. We select events where HQTot is more than 200\,PE, which removes low charge events susceptible to large systematic uncertainties and poorly reconstructed observables. These quality cuts cumulatively reduce the atmospheric muon rate from 3\,kHz to 0.1\,Hz.

\subsection{\label{sec:STV} Starting track veto}
We now define the starting track veto (STV) as diagrammed in Fig. \ref{fig:startingtrack} (first proposed in reference \cite{KyleThesis}). A veto region is constructed for each event taking into account the location of the first in-time observed photon and the expected light emission profile of an incoming muon track. This dynamic veto allows us to retain a good efficiency for astrophysical neutrinos towards lower energies while still significantly reducing the atmospheric muon and neutrino rate.

\begin{figure*}[t!]
\includegraphics[width=0.7\textwidth]{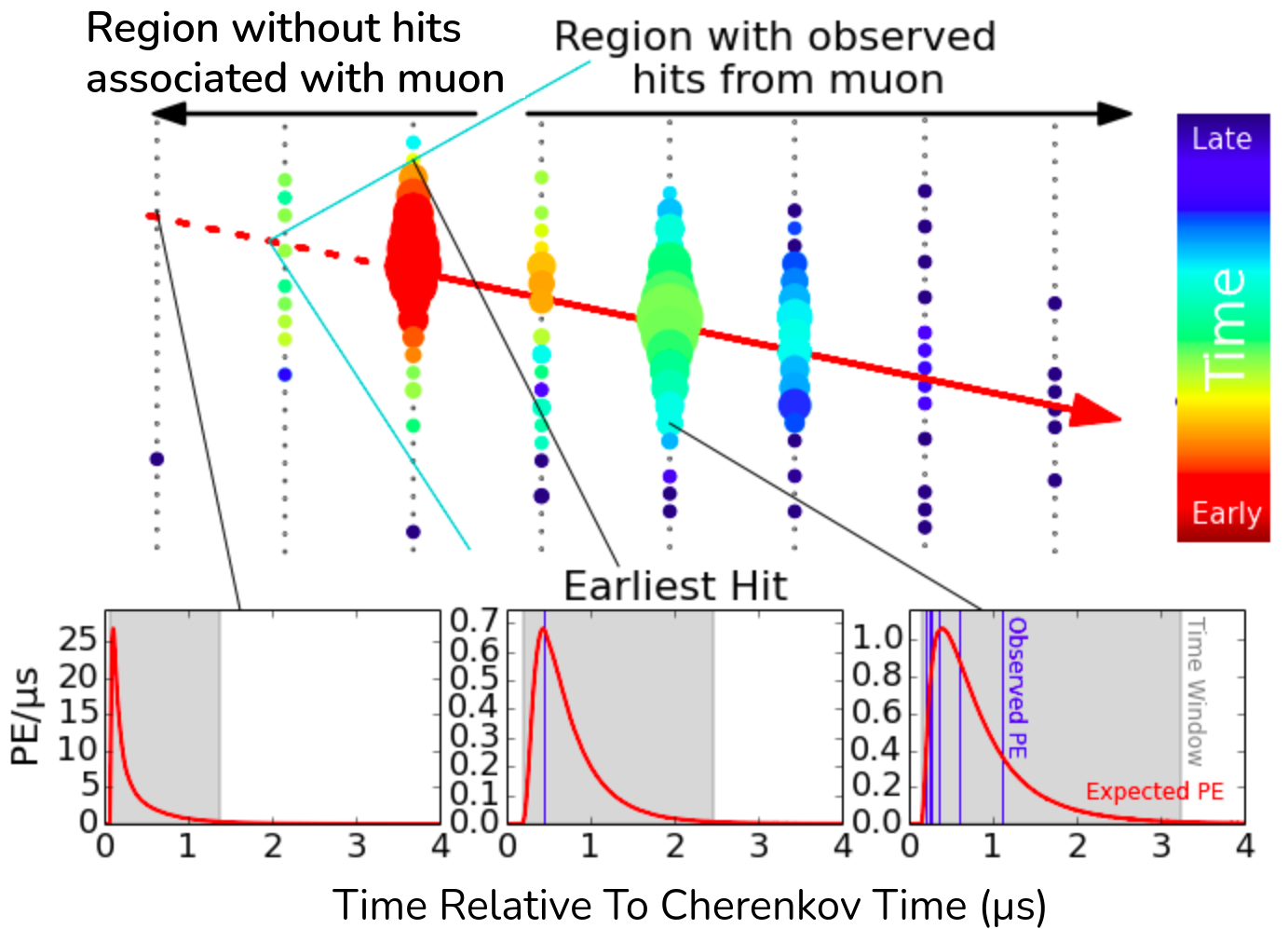}
\caption{Diagram of a starting track event, the dashed red line shows the incoming neutrino and the solid red line shows the outgoing muon track. In the lower panel, the PE probability distributions for 3 reference DOMs are shown in red (assuming the muon track emits all light) and the observed hits are show as purple vertical lines. The earliest possible hit from a muon track is then connected to the track along the Cherenkov cone direction and used to define the ``veto-region". The inferred direct Cherenkov emission cone is delineated by the cyan lines. The gray shaded regions in the sub-figures denote the time-windows corresponding to 90\% of the expected PEs. We see there are hits in the veto region that occur at much later times. These PEs, likely to be from the hadronic shower development, are excluded due to this time-window criteria. }
\label{fig:STVVertex}
\end{figure*}

The SplineMPE reconstructed muon track \cite{splinempe} is used to identify the position, direction, and time of the expected muon track. Each DOM is then assigned a probability distribution of PEs that would be expected from this muon track. This is shown as the red curves (Expected PE) in the lower panels of Fig. \ref{fig:STVVertex} and the 90\% PE expectation is shown as a gray time window. We find the first PE in the event that can be produced by this muon track (earliest in-time hit) within the allowable time windows. Using the track direction, earliest in-time hit position, and Cherenkov cone geometry, we define the ``veto-region" as the region where light should be observed assuming this particular light profile is from an incoming muon track. In the case of an actual incoming muon track, the probability of observing light in this veto-region is high. If the first hit is observed in an outer layer DOM, the event is also likely to be marked as an incoming muon. Meanwhile, the probability of observing light in this veto-region is low for a starting track event. Fig. \ref{fig:STVVertex} shows three reference DOMs for a starting track event: one before the neutrino interaction takes place, one at the location of the first observed PE, and one with multiple PEs observed. There are PEs observed in the veto region at later times from scattered photons from the hadronic shower development (or from noise), but these hits would not be included in the time window since they occur much later than hits predicted from a muon track. 

\begin{figure}[t!]
\includegraphics[width=0.45\textwidth]{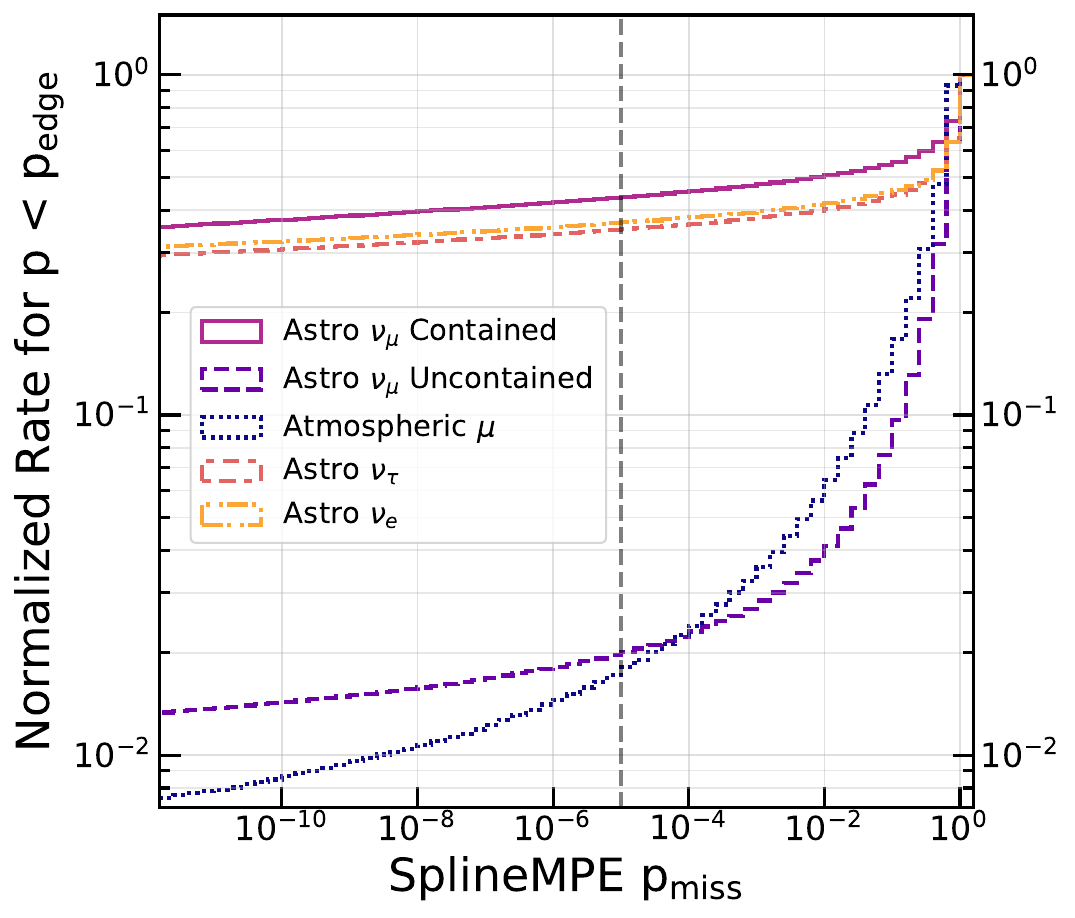}
\caption{p$_{\mathrm{miss}}$ distribution for atmospheric muons and astrophysical neutrinos. A cut requiring events to have p$_{\mathrm{miss}} < 10^{-5}$reduces atmospheric muon rates by three orders of magnitude while retaining a high signal efficiency for muon neutrino events with interaction vertices contained inside the detector volume. This figure is normalized to show p$_{\mathrm{miss}}$ shape differences between the interaction types.}
\label{fig:pmiss}
\end{figure}

The probability that a DOM detects a photon is modeled using a Poisson probability where the expected number of PEs ($\lambda$) follow the assumption that a through-going muon track actually emitted Cherenkov light. The expected number of these photons at each DOM is calculated using position, direction, timing, and energy of the particle that emitted them (e.g. higher energy particles emit more photons, particles near a DOM have a greater chance of detection). The likelihood is now defined as the product of all Poisson probabilities for all DOMs in the veto-region (VR-DOMs). This likelihood, p$_{\mathrm{miss}}$, is defined as: 
\begin{equation}
\mathrm{p}_{\mathrm{miss}}(\mathrm{k}=0, \lambda_i) = \Pi^{\text{VR-DOMs}}_{i} \frac{{e^{ - \lambda_i } \lambda_i ^k }}{{k!}}
\label{eq:pmiss}
\end{equation}
with the performance shown in Fig. \ref{fig:pmiss}. p$_{\mathrm{miss}}$ is a measurement of the probability that VR-DOMs saw no light assuming the event was throughgoing; therefore we set k = 0 for all VR-DOMs. p$_{\mathrm{miss}}$ values closer to 0 are defined as more ``starting-like" while values closer to 1 are defined as more ``throughgoing-like," events that have their interaction vertex outside of the detector's volume. We note that the presence of noise hits is negligible as the noise hit would need to occur in-time with the track to affect the placement of the vertex. Noise hits can enter the veto region by chance but the number of DOMs in the veto region is large enough such that this is a negligible effect. A cut of  p$_{\mathrm{miss}}$ = $10^{-5}$ was found to be optimal to reduce the incoming muon rates by 2 orders of magnitude while removing very little starting astrophysical neutrino events.  In addition to defining a p$_{\mathrm{miss}}$, we can also use the reconstructed vertex and direction of the muon track to estimate the length of the muon track inside the detector. A cut of 300\,m is applied to ensure the track is of sufficient length to predict its energy and direction.

Poorly reconstructed muons can pass this veto criteria if the track enters through a corridor in the detector. These corridors are defined as regions of un-instrumented detector, passing directly between adjacent rows or vertical columns of instrumented strings due to the detector's hexagonal grid. To combat this, we connect the center of gravity of charges of the event to a predefined list of these corridors. These new tracks are then refit using the same SplineMPE algorithm as before. For all events, we make a list of tracks with a reconstruction likelihood value within 2\% of the maximum likelihood, p$_{\mathrm{miss}}$ and the track length are calculated for this list. Events with a track length below 300 and events with a p$_{\mathrm{miss}}$ above $10^{-5}$ are rejected. The maximum p$_{\mathrm{miss}}$ from this procedure is then used as an input to the BDT as described in Sec. \ref{sec:BDT}. 

Finally, the starting track veto is used with a more detailed muon light emission profile. Most importantly, this improves the reconstruction of the vertex position. We then take the same set of tracks that were selected from the corridor scan and now compute the p$_{\mathrm{miss}}$ for each track, only keeping the maximum p$_{\mathrm{miss}}$ from this list of tracks and this muon light emission profile. The maximum p$_{\mathrm{miss}}$ is not used as a cut, however this is saved for use in the BDT as described in Sec. \ref{sec:BDT}. 

\subsection{\label{sec:BDT} Starting tracks boosted decision tree}
After the initial quality and veto cuts, the atmospheric muon rates are still $\sim$4 orders of magnitude higher than the expected astrophysical neutrino rate. We have removed atmospheric muons that deposit light close to the detector edge but there is still a significant number of difficult-to-detect muons remaining. To reduce this background, we use the XGBoost boosted decision tree (BDT) algorithm \cite{Chen:2016:XST:2939672.2939785} to classify events as atmospheric muons or starting muon-neutrino charged current events. We use a simulated dataset with 1 million atmospheric muons and 200,000 starting muon-neutrino charged current events to train the BDT. The dataset was split into 70/30 training/validation where the training set was used to train the BDT and the validation set was only used to select the optimal model after hyper-parameter and input variable optimization. An independent MC set with over 900,000 events was later used for the measurement. The thirteen variables used are shown in Tab. \ref{tab:BDTinputs} sorted by importance after training. 

\begin{table}[h]
\setlength{\tabcolsep}{6pt} 
\renewcommand{\arraystretch}{1} 
\begin{tabular}{l l} 
 \hline
 \hline
 Importance & Description \\ [0.5ex] 
 \hline
 1 & Number of Millipede Losses $>$ 5 GeV  \\ 
 2 & Fraction of Energy in First 10m of Track  \\
 3 & Max p$_{\mathrm{miss}}$ from simple muon hypothesis  \\
 4 & Classifier  \\
 5 & Deposited Energy  \\ 
 6 & Reconstructed Zenith  \\ 
 7 & Fraction of Hits on Outer \\ & Layer of Detector  \\ 
 8 & Distance to Detector Edge \\ & from Perpendicular to Track  \\ 
 9 & Distance to Detector Edge \\ & from Vertex Position  \\ 
 10 & Entry Position of Track - Z position  \\ 
 11 & Track Length  \\ 
 12 & Fraction of Hits within 100m \\ & Cylinder Centered at the Track \\ 
 13 & Max P$_{\mathrm{miss}}$ from detailed muon hypothesis  \\ 
 \hline
 \hline
\end{tabular}
\caption{The 13 BDT inputs sorted by importance after training. This was calculated assuming the best-fit 2-year MESE astrophysical and atmospheric neutrinos fluxes \cite{IceCube:2014rwe}.}
\label{tab:BDTinputs}
\end{table}

Atmospheric muons with zenith $(\theta) > 80^\circ $ are almost all poorly reconstructed events and the characteristics of such events are greatly different from atmospheric muons with  $\theta < 80^\circ $ which tend to be difficult-to-detect muons. The number of atmospheric muons expected greatly differs by angle; we therefore use the Precision-Recall Area Under the Curve evaluation metric \cite{10.1145/1143844.1143874} to optimize the BDT. After training, a single BDT model is used but with different cuts on BDT score for each hemisphere ($\theta > 80^\circ $ and $\theta < 80^\circ$). Using the best-fit single power law flux parameters as described in Tab. \ref{tab:systs}, we show the sorted BDT features in Tab. \ref{tab:BDTinputs}. The performance of the BDT using 1 year of IceCube data was shown in references \cite{IceCube:2021ctg,ManuelThesis}. 

\begin{figure}[t!]
\includegraphics[width=0.45\textwidth]{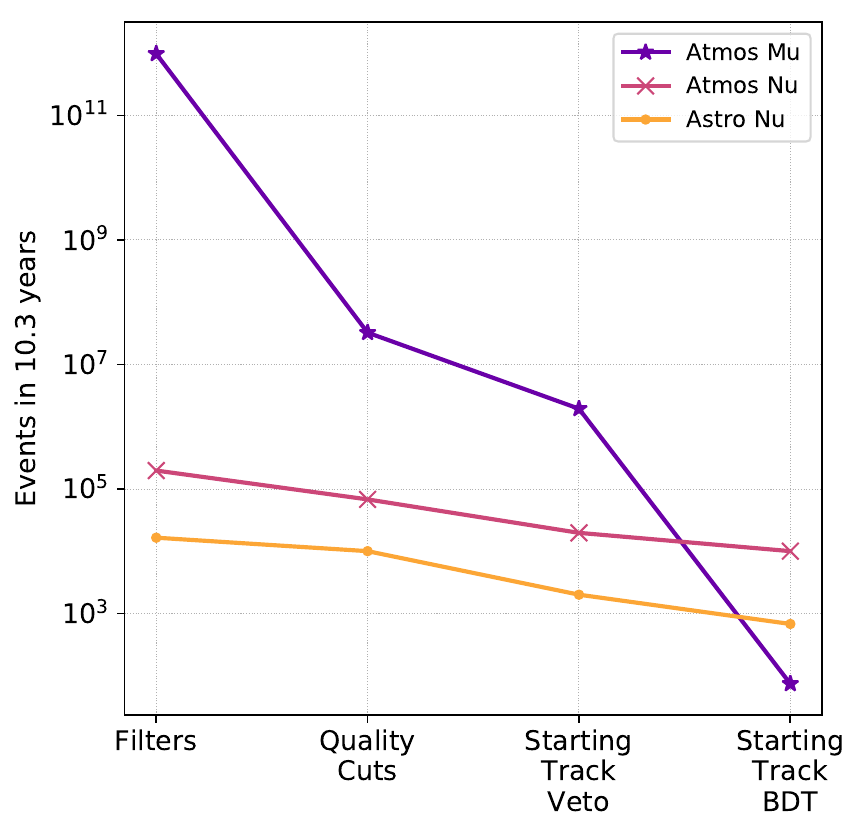}
\caption{Summary of all cuts as described in Sec. \ref{sec:EventSelection}. We start with an overwhelming atmospheric muon rate where the muons outnumber the neutrinos by at least 5 orders of magnitude. After applying all the cuts, the neutrinos outnumber the muons by 2 orders of magnitude. The MC expectation shown is pre-fit and shown for illustrative purposes only.}
\label{fig:cutssummary}
\end{figure}

Figure \ref{fig:cutssummary} summarizes the event rates from cut to cut. We see significant decrease in muon rates at each cut at the cost of some neutrino events. After applying all cuts, the muons make up $<$ 1\% of the final event rate. 10798 data events are observed between 1 TeV and 10 PeV over the entire sky. 
Table \ref{tab:eventcounts} summarizes the observed data rates compared to the rates as computed from the MC using the best-fit parameters from Tab. \ref{tab:systs}.

\begin{table}[h]
\begin{tabular}{l | l | l} 
\hline
\hline
& Events $\theta < 80^\circ$ & All Events \\
\hline
Astro Nu & 298 & 680 \\ 
Atmos Conv. Nu & 980 & 10042 \\ 
Atmos Conv. Mu & 42 & 75 \\ 
Total MC & 1320 & 10797 \\
Data & 1365 & 10798 \\
\hline
\hline
\end{tabular}
\caption{The observed data events and fitted MC events broken up by type. The left column has an additional cut on zenith $< 80^\circ$ to emphasize the higher proportion of astrophysical neutrinos in the southern sky.}
\label{tab:eventcounts}
\end{table}

\section{\label{sec:measurement} Measurement Methodology}
\subsection{\label{sec:Stats} Statistical analysis}
The measurement of the diffuse flux utilizes the forward folding likelihood technique. All simulated and observed data events are placed into two-dimensional binning as summarized in Tab. \ref{tab:bins}, totaling 190 bins. Each bin has a corresponding expectation value as computed using simulated data. The simulated data is a sum of the astrophysical neutrinos, atmospheric conventional neutrinos, atmospheric prompt neutrinos, and atmospheric muons: 
\begin{equation}
\lambda = \lambda_{Astro} + \lambda_{Conv} + \lambda_{Prompt} + \lambda_{Muon}.
\label{eq:lambda}
\end{equation}

\begin{table}[h]
\begin{tabular}{||l l l||} 
 \hline
 Observable & Bin Range & Number of Bins \\ [0.5ex] 
 \hline\hline
 \multirow{2}{*}{Energy} & \multirow{2}{*}{1 TeV to 1 PeV} & 18 (log) \\
 && [+1 overflow]\\ 
 Cosine Zenith & -1 to 1 & 10 (linear) \\ 
 \hline
\end{tabular}
\caption{Summary of 190 bins used to define observed and expected number of events in the likelihood Eq. \ref{eq:llh}. One overflow bin is used to capture events from 1 PeV to 10 PeV.}
\label{tab:bins}
\end{table}

Each of these terms is modified according to a flux normalization (atmospheric component and astrophysical component) and spectral index (astrophysical component only).

We now define the probability of having observed $k$ events while expecting $\lambda$-events using a Poisson probability. The likelihood function is defined as the product of all 190 Poisson probabilities. To take into account systematic uncertainties (nuisance parameters), described in greater detail in Section \ref{sec:SystematicUncertainties}, we introduce a modification to our expectation value $\lambda = \lambda (\eta, \epsilon$). The set of $\eta,\epsilon$ are summarized in table \ref{tab:systs}. The $\epsilon$ parameters are aided by external measurements using a Gaussian function $(\mathcal{N})$ with a mean $(\mu)$ and standard deviation $(\sigma)$ to constrain the likelihood. Nuisance parameters without external constraints are modeled using a uniform distribution ($\eta$). The modified likelihood function including these modifications is:
\begin{equation}
\mathcal{L}(\lambda(\eta,\epsilon) | k) = \Pi ^{190}_{i=1} (\frac{{e^{ - \lambda_i } \lambda ^k }}{{k!}}) \cdot \Pi^{6}_{j=1} \frac{e^{-(\epsilon_j-\mu_j)^{2}/2\sigma_{j}^{2}}}{\sigma_{j} \sqrt{2\pi}}.
\label{eq:llh}
\end{equation}

To compute the parameters $\hat{\Theta}$ that best describe the data, we run a minimization of the negative log of the likelihood function using the Minuit2 C++ library \cite{James:1975dr,iminuit}. To compute the confidence intervals on the best fit set of parameters, we use the profile likelihood technique. The likelihood is now defined as the likelihood with the set of parameters $\Theta$ such that the likelihood function is once again maximized. We define the test-statistic for this measurement as the negative log likelihood ratio of this likelihood function at $\Theta$ with respect to the likelihood function at $\hat\Theta$. This test-statistic is:. 
\begin{equation}
\mathcal{TS}(\Theta) = -2 \mathrm{log} (\frac{\mathcal{L}(\lambda(\Theta) | k)}{\mathcal{L}(\lambda(\hat\Theta) | k)} ).
\label{eq:ts}
\end{equation}

The confidence intervals are then presented using Wilks' theorem \cite{Wilks}. 

\subsection{\label{sec:SystematicUncertainties}  Systematic uncertainties}
The ESTES data selection is a significant increase in the total number of events as compared to previous starting event event selections. This large increase motivated an expanded treatment of systematic uncertainties. This section describes all systematic parameters used in this measurement. A summary of all parameters (pre)post-fit is available in Tab. \ref{tab:systs}. The last column shows the best-fit points after the measurement is performed with 68\% confidence intervals as defined in Section \ref{sec:Stats}. When relevant, the $\pm1\sigma$ constraints are shown as 2D templates in Appendix \ref{sec:DetectorSystematicPriors}.

\begin{table*}[t!]
\setlength{\tabcolsep}{6pt} 
\renewcommand{\arraystretch}{1.5} 
\begin{tabular}{c c c c c} 
 \hline
 \hline
 Parameter & Boundary & Constraint ($\mu \pm \sigma$) & Best-Fit & Description \\ [0.5ex] 
 \hline
 \multicolumn{5}{>{\bfseries}p{\textwidth}}{Astrophysical Flux Parameters}\\
 $\Phi_\mathrm{Astro}$ & $[0,\infty)$ & - & $ 1.68 ^{+0.19}_{-0.22}$ & Astrophysical neutrino flux normalization\\ 
 $\gamma_\mathrm{Astro}$ & $[0,\infty)$ & - & $ 2.58 ^{+0.10}_{-0.09}$ & Astrophysical neutrino flux spectral index \\
 \hline
 \multicolumn{5}{>{\bfseries}p{\textwidth}}{Atmospheric Flux Parameters}\\
 $\Phi_\mathrm{muon}$ & $[0,\infty)$ & - & $ 0.6 \pm 0.4$ & Atmospheric muon flux normalization \\ 
 $\Phi_\mathrm{conv}$ & $[0,\infty)$ & - & $ 1.5 \pm 0.3$ & Atmospheric conventional neutrino flux normalization\\ 
 $\Phi_\mathrm{prompt}$ & $[0,\infty)$ & - & $ < 3.19 $ (90\% U.L.) & Atmospheric prompt neutrino flux normalization\\
 $\epsilon_{\nu\bar\nu\text{-}\mathrm{ratio}}$ & [0,2] & 1 $\pm$ 0.10 & $ 1.04 \pm 0.08$ & $\nu\bar\nu$-ratio \\
 $\eta_{\mathrm{H4a-GST}}$ & [$-2$,+1] & - & $ -1.4 \pm 0.4$ & H4a-GST cosmic ray flux model interpolation \\
 $\eta_{\mathrm{2.3c-DPMJet}}$ & [$-2$,+1] & - & $ -0.6 \pm 0.6$ & 2.3c-DPMJet hadronic interaction model interpolation  \\
 $\eta_{\mathrm{Self-Veto}}$ & [1, 3] & - & $ 2.1 ^{+0.1}_{-0.3}$ & Self-veto muon rejection intensity, $log_{10}(\frac{\mathrm{Energy}}{1  \mathrm{GeV}})$ units  \\
 \hline
 \multicolumn{5}{>{\bfseries}p{\textwidth}}{Detector Systematic Parameters}\\
 $\epsilon_\mathrm{Scattering}$ & [0.8,1.2] & 1 $\pm$ 0.05 & $ 1.04 \pm 0.03$ & Bulk-ice model scattering coefficient scaling \\ 
 $\epsilon_\mathrm{Absorption}$ & [0.8,1.2] & 1 $\pm$ 0.05 & $ 0.98 \pm 0.03$ & Bulk-ice model absorption coefficient scaling  \\ 
 $\epsilon_\mathrm{Angular, DOM (p_{0})}$ & [$-0.5$,0.3] & $-0.3 \pm 0.5$ & $ -0.3 \pm 0.3$ & Angular PM acceptance parameter p0\\ 
 $\epsilon_\mathrm{Angular, DOM (p_{1})}$ & [$-0.10$,0.05] & $-0.04 \pm 0.10$ & $ -0.09 \pm 0.05$ &  Angular PM acceptance parameter p1\\ 
 $\epsilon_\mathrm{Overall, DOM}$ & [0.8,1.2] & 1 $\pm$ 0.10 & $ 0.91 \pm 0.05$ & Absolute DOM acceptance\\ 
 \hline
 \hline
\end{tabular}
\caption{Summary of all parameters used in the measurement of the astrophysical diffuse flux using a single power law. All parameters with constraints are modeled as a Gaussian penalty term in the likelihood. All parameters are assumed to be independent.}
\label{tab:systs}
\end{table*}

\subsubsection{\label{sec:theorysyst} Atmospheric Flux Systematics}
The atmospheric neutrinos were modeled using the Gaisser H4a cosmic ray \cite{H4a} and Sibyll 2.3c hadronic interaction \cite{23c} models. We treat the normalization of the conventional and prompt neutrino fluxes and the conventional atmospheric muon flux as nuisance parameters by using overall normalization factors for each component. These factors are labelled $\Phi_\mathrm{conv}$, $\Phi_\mathrm{prompt}$, and $\Phi_\mathrm{muon}$. The $\epsilon_{\nu\bar\nu\text{-}\mathrm{ratio}}$ systematic uncertainty is centered at 1.0 with a Gaussian prior of 0.10. This ratio term controls the relative contributions from atmospheric neutrinos to anti-neutrinos ($\epsilon_{\nu\bar\nu\text{-}\mathrm{ratio}} = 2\nu / (\nu + \bar\nu)
$) and is used a correction term to the theoretical expectation. The choice of 10$\%$ is an estimate derived by comparing various atmospheric flux models \cite{Collin_2015} and taking the maximal differences. The same ratio term is used for conventional and prompt atmospheric neutrinos. 

\begin{figure*}[t!]
\includegraphics[width=0.49\linewidth]{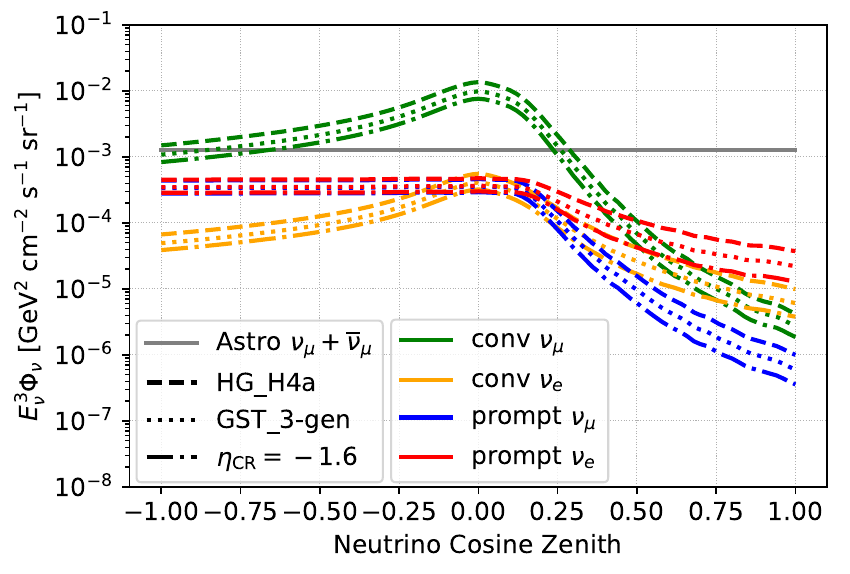}
\includegraphics[width=0.49\linewidth]{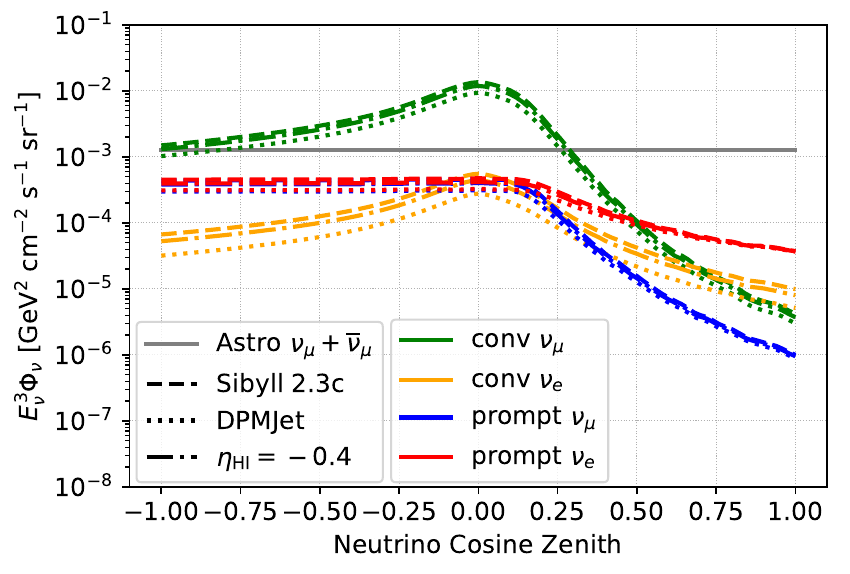} 
\caption{Conventional and prompt atmospheric neutrino fluxes at 50 TeV shown as a function of cosine zenith for the various choices of cosmic ray models (left) and hadronic interaction models (right). The fluxes labeled as $\eta_{\mathrm{CR}/\mathrm{HI}}$ are the best-fit interpolated fluxes as defined in Section \ref{sec:theorysyst}. We observe large differences between the theoretical cosmic-ray flux model and the preferred model, but minor differences between the hadronic models and the preferred model. We use the SPL best fit $\eta_{\mathrm{Self-Veto}} = 2.1$ (126 GeV) for the fluxes shown here.}
\label{fig:CRHImodels}
\end{figure*}

The $\eta{\mathrm{H4a-GST}}$ systematic uncertainty was motivated by the expected shape differences between different cosmic ray flux models parametrized in MCEq. This parameter was first introduced in a recent measurement of the flux using tracks from the northern sky\cite{IceCube:2021uhz}. When $\eta{\mathrm{H4a-GST}} = 0$, the data agrees perfectly with the H4a cosmic ray flux model and when $\eta{\mathrm{H4a-GST}} = -1$ the data prefers the GST cosmic ray flux model \cite{Gaisser:2013bla}. A linear interpolation in log-space for the difference of the predicted fluxes was used to model this uncertainty. We allowed some flexibility by constraining the predicted flux at $\pm 2$ and modeling the uncertainty as a flat prior. The expected atmospheric neutrino fluxes and the best-fit $\eta{\mathrm{H4a-GST}}$ flux are shown in Fig. \ref{fig:CRHImodels}. The same $\eta{\mathrm{H4a-GST}}$ term was used for conventional and prompt atmospheric neutrino fluxes. The $\eta{\mathrm{2.3c-DPMJet}}$ systematic uncertainty is modeled using the same technique as described for $\eta{\mathrm{H4a-GST}}$ but this time interpolating between the Sibyll 2.3c and the DPMJet hadronic interaction models \cite{Fedynitch:2015kcn}. The $\eta{\mathrm{2.3c-DPMJet}}$ was introduced as an alternative to using Barr parameters as described in \cite{PhysRevD.74.094009}.

\begin{figure}[t!]
\includegraphics[width=0.45\textwidth]{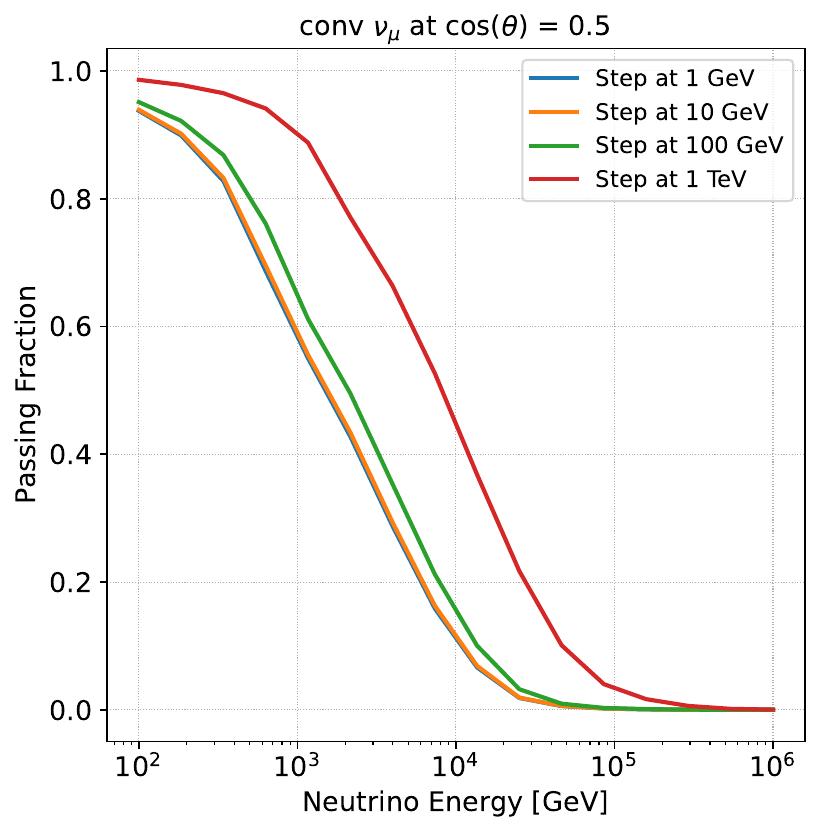}
\caption{Conventional atmospheric muon neutrino passing fractions at cosine zenith = 0.5. The different colors correspond to different choices of muon energy used for the Heaviside step function. The muon threshold energy is treated as a free parameter in our model.}
\label{fig:passingfracs}
\end{figure}

As initially described in Section \ref{sec:Flux}, for cosine zenith $> 0$, both the conventional and prompt atmospheric neutrino fluxes experience an energy, cosine zenith, and depth-dependent suppression due to the self-veto effect. In previous IceCube cascade-dominated measurements using the southern sky \cite{IceCube:2014rwe,SBUCasc}, it was assumed that muons with energy greater than 1 TeV are all rejected. This rejection probability is defined as a Heaviside step function \cite{PhysRevD.79.043009,PhysRevD.90.023009}. While the choice of 1 TeV is well motivated (muons are minimum ionizing particles below this energy), it is conservative to treat this energy threshold as a free-parameter in the measurement. The $\eta_{Self-Veto}$ nuisance parameter is defined as a parameter that weakens and strengthens the muon rejection probability function. The introduction of this nuisance parameter is motivated such that we minimize the potential bias on the flux measurement due to choice of muon rejection probability model. The value corresponds to the muon energy used in a Heaviside function. Different choices of this probability are used to calculate the atmospheric neutrino flux. Figure \ref{fig:passingfracs} shows the passing fraction ($P_{frac}$) which is the ratio of the flux with the self-veto divided by the flux without the self-veto effect. We note there are minor differences at lower muon energies, but for energies above 100 GeV large differences in $P_{frac}$ are expected. We parametrize the $\eta_{Self Veto}$ term as a function of the threshold muon energy for each energy-zenith bin used in the measurement. The preferred threshold for this data is 126 GeV as shown in Tab. \ref{tab:systs}. 

\subsubsection{\label{sec:detectorsyst} Detector systematics}
Detector uncertainties are defined as any systematic uncertainty that can affect the detector response due to the modeling of the Cherenkov photons in the simulation. These arise due to limited knowledge of the optical properties of the South Pole ice and overall PMT response. The five systematic parameters discussed in Tab. \ref{tab:systs} are parameterized by rerunning the same set of events through the detector simulation under various ice and detector configurations. A ``baseline" simulation set is centered at the mean and then varied within the allowed range to parameterize the detector response per nuisance parameter. Linear interpolation is assumed between the simulated ranges as shown in Tab. \ref{tab:systs} for each bin in the the energy/zenith observable space. The $1\sigma$ detector systematic constraints are shown in appendix \ref{sec:DetectorSystematicPriors} with respect to the baseline simulation.


The South Pole bulk-ice refers to the ice between the strings in the detector. A depth-dependent parametrization \cite{SPICE1,SPICE2} of the photon scattering \cite{Askebjer:97} and absorption \cite{Price:97} coefficients is used in simulations to account for the effect of glacial ice impurities\cite{depthice,journal_of_glaciology_2013} and structural properties of the ice, on photon propagation. We model $\epsilon_\mathrm{Scattering}$ and $\epsilon_\mathrm{Absorption}$ as Gaussian terms centered at nominal scattering/absorption parameters with a $\pm5\%$ overall uncertainty. These $\pm1\sigma$ constraints are shown in Fig. \ref{fig:scat} and Fig. \ref{fig:abs} in App. \ref{sec:DetectorSystematicPriors}.

\begin{figure}[t!]
\includegraphics[width=0.45\textwidth]{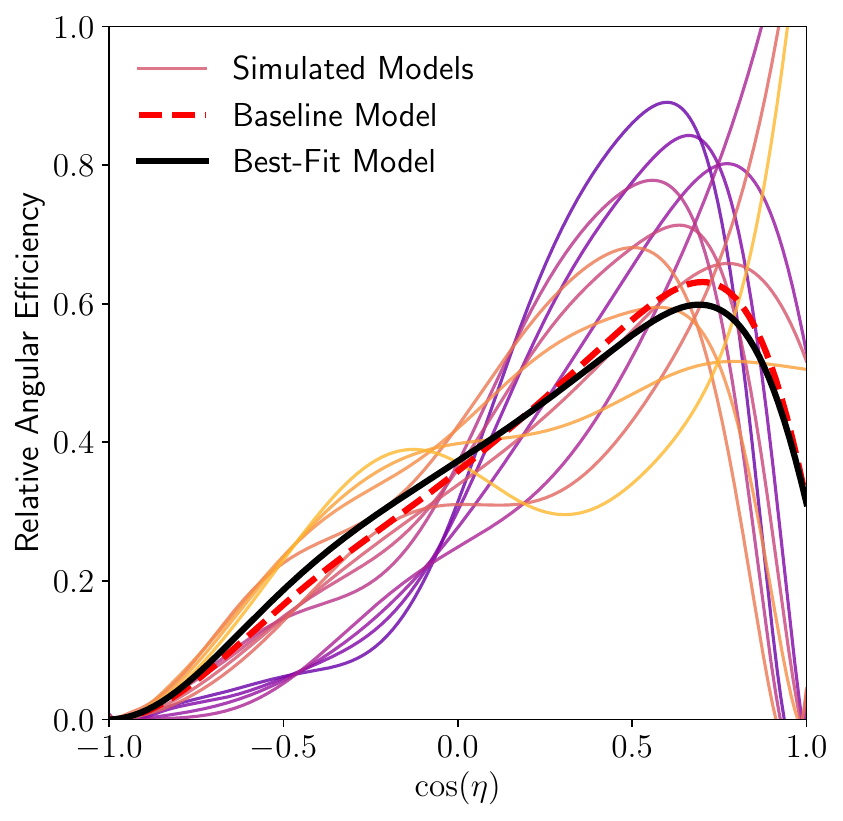}
\caption{The optical module angular response function. The discrete colors represent slices of the $p_{1}$ and the color gradients represent the range of $p_{0}$ simulated. These parameters are modeled as a continuous set of parameters in the likelihood. $\eta$ is the photon incident angle where cos $\eta$ = 1 is the photon entering upwards incident with the PMT.}
\label{fig:holeice}
\end{figure}

The photomultiplier tube in an IceCube DOM points downward, causing a large zenith angle dependence in the photon detection efficiency \cite{PhysRevD.79.062001, PhysRevD.79.062001}. Up-going photons that enter the DOM directly will enter the PMT head-on resulting in maximal photon detection efficiency, whereas down-going photons that enter the DOM need to scatter within the optical module itself or the ice surrounding it.  The columns of refrozen ice containing the DOMs have higher concentrations of impurities, particularly air bubbles \cite{holeice2016}. This results in the hole-ice having different optical properties when compared to that of the bulk-ice \cite{holeice,holeice2019}. We model the effects of the hole-ice as a single angular response function using two parameters, $\epsilon_\mathrm{Angular, DOM (p_{0})}$ and $\epsilon_\mathrm{Angular, DOM (p_{1})}$ with arbitrary units \cite{IceCubeCollaboration:2023wtb} (parameters hold no physical meaning themselves). These parameters were simulated over the ranges shown in Fig. \ref{fig:holeice} and treated as independent parameters. The colors represent discrete choices of $p_0$ and $p_1$ parameters simulated for the ranges indicated in Tab. \ref{tab:systs}. The $\pm1\sigma$ constraints are shown in Fig. \ref{fig:p0} and Fig. \ref{fig:p1} in App. \ref{sec:DetectorSystematicPriors}.

The DOM efficiency uncertainty represents the cumulative systematic error of the absolute sensitivity of the sensor within IceCube \cite{icecubepmt}. Calibration studies of the absolute sensitivity found differences between the simulated charge and observed charge from 5\% to 10\% in some regions of the detector \cite{Tosi:2014rln}. We model the DOM efficiency using a Gaussian constraint term centered at $1.0$ with an uncertainty of $\pm 0.10$ as motivated by muon studies ($\epsilon_\mathrm{Overall, DOM}$) \cite{Aartsen_2014}. This overall scaling factor is applied to all IceCube DOMs. The $\pm1\sigma$ constraints are shown in Fig. \ref{fig:domeff} in appendix \ref{sec:DetectorSystematicPriors}.

\section{\label{sec:Results} Diffuse Flux Measurement}
\subsection{\label{sec:Measurement} Measurement of the diffuse flux assuming a single power law flux}
A search for the astrophysical neutrino flux is first performed under an isotropic single power law flux (SPL) hypothesis, and the best-fit SPL parameters are determined to be: 
\begin{equation}
\begin{split}
&\Phi_{Astro}^{Total} = \phi_{\mathrm{Astro}}^{\mathrm{per-flavor}} \times (\frac{\mathrm{E}_{\nu}}{100 \mathrm{TeV}})^{-\gamma} \times \mathrm{C}_{0}, \\
&\mathrm{C}_{0} = 3 \times 10^{-18} \times \mathrm{GeV}^{-1} \mathrm{cm}^{-2} \mathrm{s}^{-1}  \mathrm{sr}^{-1} \\
&\mathrm{where},\\
&\phi_{\mathrm{Astro}}^{\mathrm{per-flavor}} = 1.68 ^{+0.19}_{-0.22}, \quad \gamma = 2.58 ^{+0.10}_{-0.09}\\
\label{eq:SPL}
\end{split}
\end{equation}

This model (and all following models) assume $\nu_{e}:\nu_{\mu}:\nu_{\tau} = 1:1:1$ and $\nu:\bar\nu=1:1$ arriving at the surface of the Earth. In Eq. \ref{eq:SPL}, $\phi_{\mathrm{Astro}}^{\mathrm{per\ flavor}}$ refers to the per flavor normalization ($\nu+\bar\nu$) and is defined as a unit-less number. We introduce $C_{0}$ as a constant that carries the units for the diffuse flux and a factor of 3 to compensate for the three flavors. Unless explicitly defined otherwise, whenever we refer to the astrophysical normalization we are referring to the per-flavor normalization.

\begin{figure}[t!]
\includegraphics[width=0.49\textwidth]{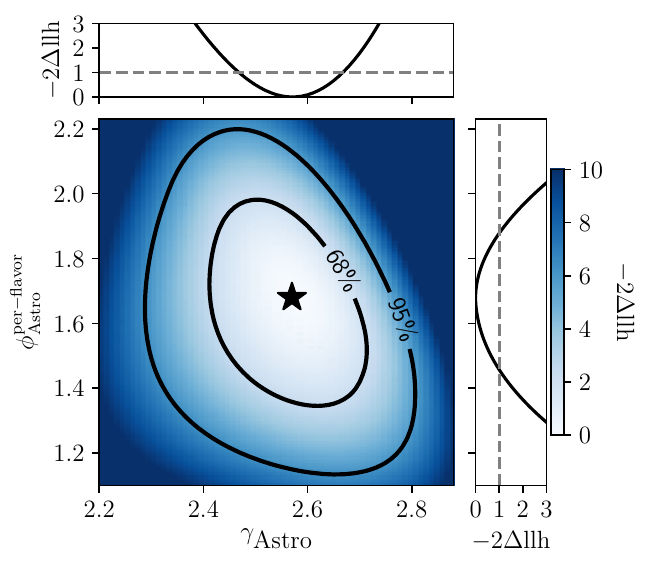}
\caption{Single power law flux likelihood scan shown in 1D and 2D. Wilks' theorem is used to define the 68\% and 95\% confidence intervals. The best-fit normalization and spectral index is shown as a black star.}
\label{fig:spl6895}
\end{figure}

\begin{figure*}[t!]
\includegraphics[width=0.48\textwidth]{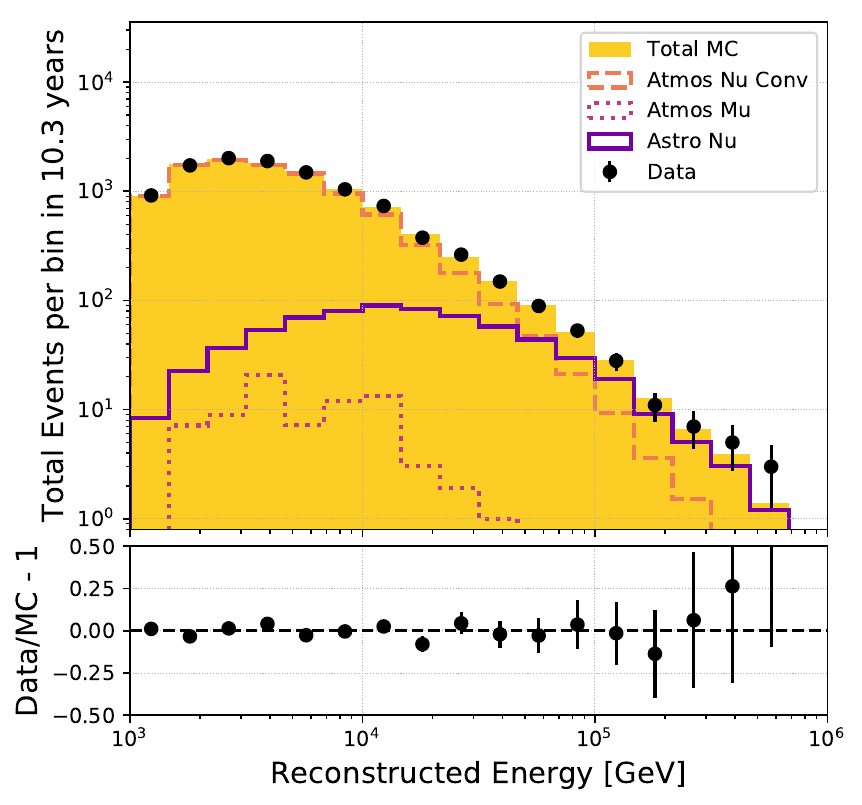}
\includegraphics[width=0.48\textwidth]{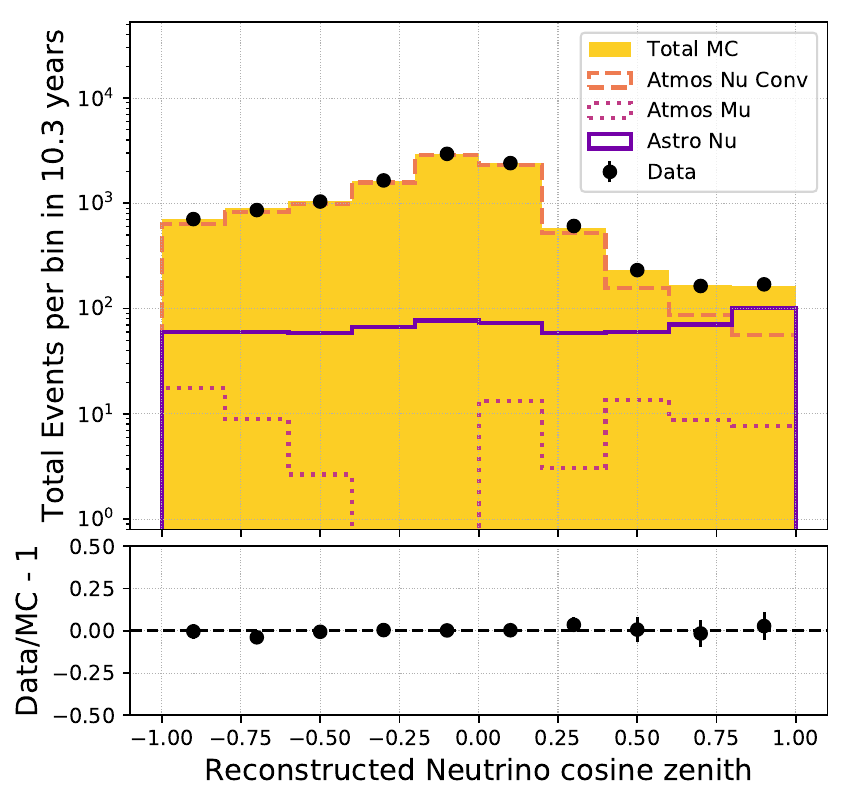}
\caption{The reconstructed energy and cosine zenith distributions for data and simulated data using the best-fit parameters from the single power law flux measurement. The astrophysical neutrinos are shown as a solid purple line, the atmospheric neutrino and muon expectations are shown as dashed and dotted lines respectively. The error bars shown for data are due to Poisson statistics only.}
\label{fig:datamc}
\end{figure*}

\begin{figure}[t!]
\includegraphics[width=0.45\textwidth]{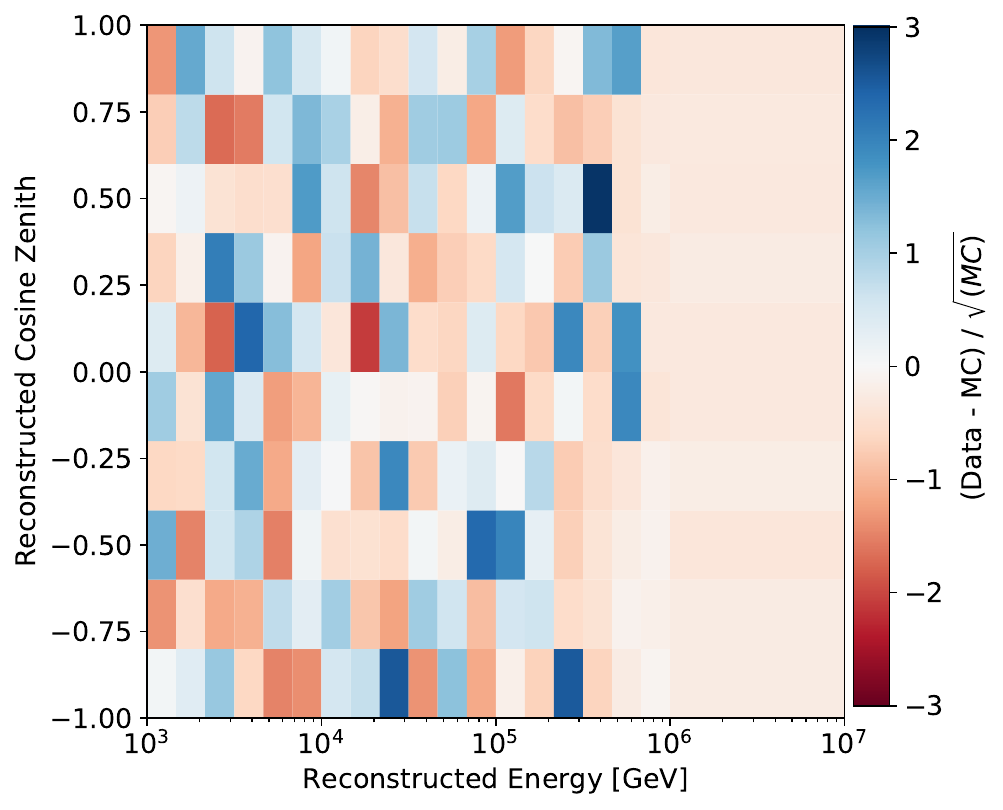}
\caption{The pulls on each energy/zenith bin for the data and Monte Carlo. No unexpected effects are observed.}
\label{fig:pullplot}
\end{figure}

All of the parameters and their 1$\sigma$ confidence intervals are shown in Tab. \ref{tab:systs}. The two-dimensional confidence intervals for the two parameters of interest are shown in Fig. \ref{fig:spl6895} using the profile likelihood assuming Wilks' theorem. A comparison of the 68\% confidence intervals to the most recent IceCube results is shown in Fig. \ref{fig:splsummary} and discussed in Section \ref{sec:IceCubeSummary}. Using the the best-fit parameters, we now compare the simulated data to the observed data in Fig. \ref{fig:datamc} for energy and cosine zenith distributions of the events. A goodness of fit test with a p-value = 0.7 using the saturated Poisson likelihood test \cite{BAKER1984437} confirms excellent agreement between the data and simulation. We also compute the pulls on the data and show them in Fig. \ref{fig:pullplot} over the full 2D observable space. There are no significant deviations observed from our expectation.

\begin{figure}[t!]
\includegraphics[width=0.45\textwidth]{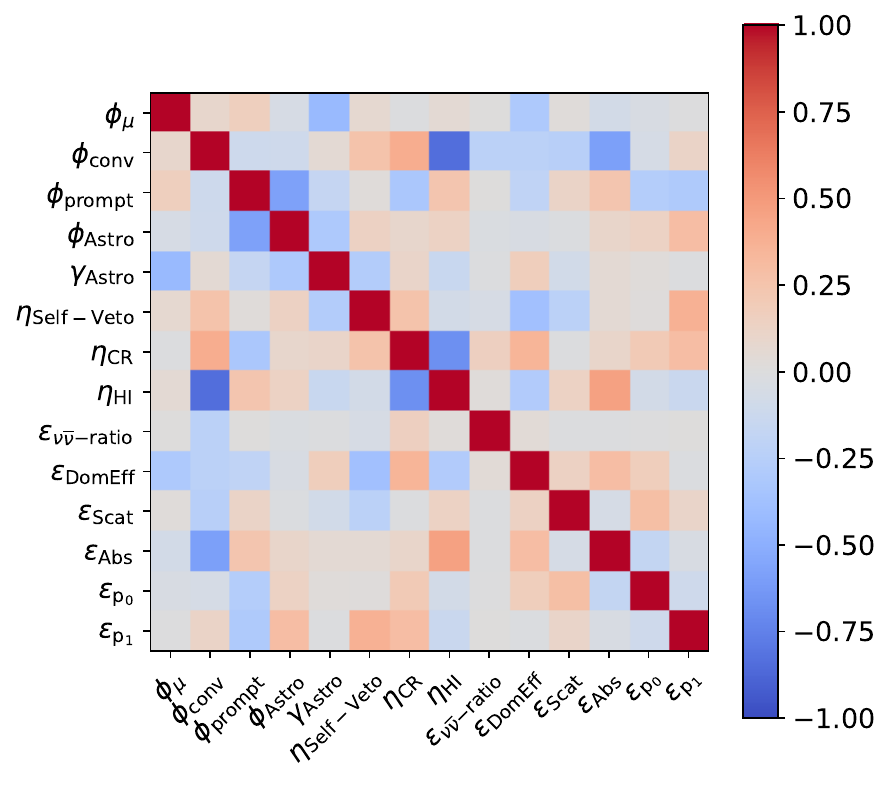}
\caption{Correlation matrix for the single power law flux measurement for all parameters used in the likelihood fit. The correlations are computed using the Hessian matrix.}
\label{fig:correlationmatrix}
\end{figure}

\begin{figure}[t!]
\includegraphics[width=0.49\textwidth]{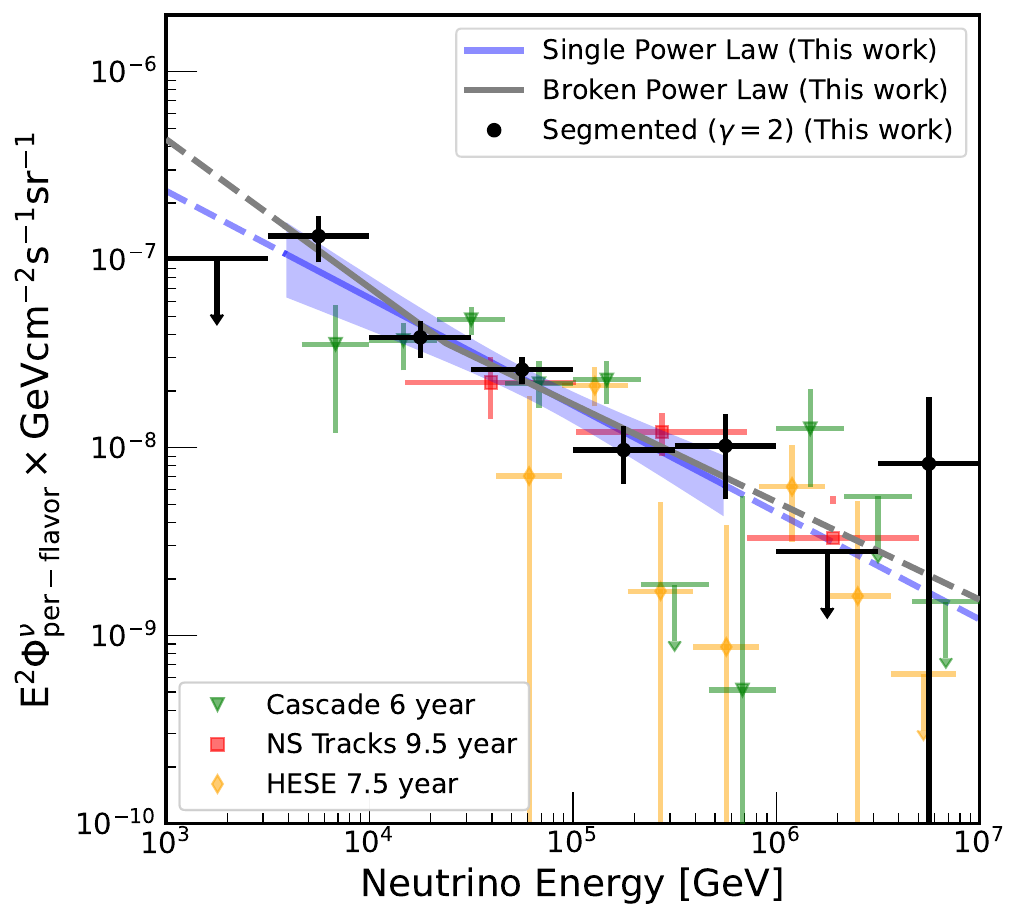}
\caption{The per flavor astrophysical neutrino flux shown as a function of energy. The black points are the segmented power law flux measurement assuming a spectral index of -2. The blue line with error bands corresponds to the SPL measurement as shown in Fig. \ref{fig:spl6895}. The blue shaded region is the 90\% sensitive energy range. The gray line is a fit to data assuming a broken power law flux. We include results from recent IceCube publications for direct comparison \cite{SBUCasc,IceCube:2021uhz,HESENew}.}
\label{fig:fluxsummary}
\end{figure}

The 90\% sensitive energy range for the astrophysical flux model is 3-550 TeV using the techniques described in \cite{Schoenen:696221}, reaching lower energies than previous diffuse analyses in IceCube \cite{SBUCasc,IceCube:2021uhz,HESENew}. We show this energy range as a blue shaded region and solid lines in Fig. \ref{fig:fluxsummary}. We remind the reader that the previous lowest energy flux measurement was dominated by cascades (electron and tau neutrinos), a dedicated discussion of these differences is in Sec. \ref{sec:IceCubeSummary}. Despite these differences in sensitive energy ranges, the previous IceCube single power law flux measurements are consistent with this measurement, 

The correlation matrix between all physics and nuisance parameters is shown in Fig. \ref{fig:correlationmatrix} computed using the Hessian matrix \cite{Fisher1,Fisher2,Fisher3}. The spectral index and astrophysical normalization are not correlated or anti-correlated with any particular parameter. The strongest correlation for astrophysical normalization is with the hadronic interaction model uncertainty, whereas the strongest correlation for spectral index is with the atmospheric muon flux.

\subsection{\label{sec:SegmentedResults} Measurement of the diffuse flux assuming a segmented power law}

We now characterize the astrophysical flux with an isotropic, segmented power law over 1 TeV-100 PeV, defined as a step-function of single power law fluxes fixed at $\gamma = 2$, with two measured bins per energy-decade, as shown in Eq. \ref{eq:segmented}:
\begin{equation}
\Phi_{Astro}^{Total} = \Sigma_{i=1}^{8} \phi_{i}  \times (\frac{\mathrm{E}_{\nu,i}}{100 \mathrm{TeV}})^{-2} \times C_{0}.
\label{eq:segmented}
\end{equation}
The measured results, $\mathrm{E}_{\nu,i}$ and $\phi_{i}$ ($\nu:\bar\nu$), are defined in Tab. \ref{tab:segmentedresults}. The width of each bin was chosen such that we minimized bin-to-bin correlations. This measurement was performed over the entire sky with all nuisance parameters from Tab. \ref{tab:systs}, and it allows us to quantify energy dependent effects on the flux in a model-independent way.

\renewcommand{\arraystretch}{1.5} 
\begin{table}[h]
\begin{tabular}{@{\hspace{5pt}}c @{\hspace{10pt}}c @{\hspace{10pt}}l @{\hspace{8pt}}c @{\hspace{8pt}}c} 
 \hline\hline
 Bin$_i$ & Energy$_{\nu,i}$ & Energy Range & $\phi_i$ ($\pm 1 \sigma$) & $\frac{\phi_{i, GP} - \phi_i}{\phi_i}$ \\ 
 \hline
 1 & 1.78 TeV & [1 - 3.16 TeV] & $ 0.0 ^{+10.2}_{-0}$ & 0\%\\ 
 2 & 5.62 TeV & [3.16 - 10 TeV] & $ 13.3 ^{3.67}_{-3.67}$ & -3.99\% \\ 
 3 & 17.8 TeV & [10 - 31.6 TeV] & $ 3.86 ^{+0.85}_{-0.85}$ & -14.44\% \\ 
 4 & 56.2 TeV & [31.6 - 100 TeV] & $ 2.60 ^{+0.43}_{-0.43}$ & -7.57\% \\ 
 5 & 178 TeV & [100 - 316 TeV] & $ 0.97 ^{+0.33}_{0.33}$ & -7.22\% \\ 
 6 & 562 TeV & [316 TeV - 1 PeV] & $ 1.02 ^{0.49}_{-0.49}$ & -1.96\% \\ 
 7 & 1.78 PeV & [1 - 3.16 PeV] & $ 0.00 ^{+0.28}_{-0}$ & 0\% \\ 
 8 & 5.62 PeV & [3.16 - 10 PeV] & $ 0.82 ^{+1.04}_{-0.82}$ & -26.83\% \\ 
 \hline
 \hline
\end{tabular}
\caption{The results of the segmented power law fit as show in Fig. \ref{fig:fluxsummary}. All normalization components are fit simultaneously including all systematic uncertainties from Tab. \ref{tab:systs}. The uncertainties are the 68\% confidence intervals assuming Wilks' thereom. The rightmost column compares the segmented power law fit with a refit done using a galactic plane Gaussian prior term described in Sec. \ref{sec:MeasurementHemisphere}.}
\label{tab:segmentedresults}
\end{table}

For each $\phi_i$, a range of neutrino energies is used. When plotting each normalization in Fig. \ref{fig:fluxsummary}, the median energy for these energy ranges in log-space is used to compute the total astrophysical flux per flavor. When the best-fit $\phi_{i} = 0$, a 68$\%$ upper limit is quoted. All segments are consistent with the single power law flux measurement, indicating a lack of evidence for energy dependent structure beyond a single power law. Previous IceCube measurements are shown for direct comparison \cite{SBUCasc,IceCube:2021uhz,HESENew}, and they also did not find any evidence beyond the single power law. We note each dataset used different bins for their analysis given their various strengths and weakness, further discussed in Sec. \ref{sec:IceCubeSummary}. An analysis of IceCube cascade events \cite{SBUCasc} found hints of a hardening of the flux towards lower energies but we do not observe this hardening in this sample. The compatibility of the data samples is discussed in greater detail in Sec. \ref{sec:IceCubeSummary}.

At the highest energies, a non-zero flux was observed from 3-10 PeV. This measurement is consistent with the Glashow Resonance (GR) \cite{PhysRev.118.316} flux measurement from IceCube \cite{IceCube:2021rpz}. Monte Carlo only studies found the most likely GR event topology is from $\bar\nu_{e} + e \rightarrow W \rightarrow \mu + \nu_{\mu}$ or $\bar\nu_{e} + e \rightarrow W \rightarrow \tau + \nu_{\tau}$ where the $\tau$ decays leptonically $\tau \rightarrow \nu_{\tau} + \mu + \nu_{\mu}$. This starting track would have no hadronic shower but would still contain an energetic muon track \cite{Bhattacharya:2011qu}. The resulting muon would only carry about 100 - 500 TeV of the initial neutrino energy preventing us from identifying the single data event using the data sample as presented in this work. 

\subsection{\label{sec:MeasurementBPL} Measurement of the diffuse flux assuming a broken power law}
We now characterise the astrophysical flux with an isotropic, broken power law (BPL), 
\begin{equation}
\begin{split}
&\Phi_{Astro}^{Total} = \phi_{0} \times (\frac{\mathrm{E}_{\mathrm{break}}}{100 \mathrm{TeV}})^{-\gamma_2} \times \mathrm{C}_{0},
\\
&\phi_{0} = 
\left\{
    \begin{array}{c}
        \phi^{\mathrm{Astro}}_{\mathrm{per-flavor}}  \times (\frac{\mathrm{E}}{\mathrm{E}_{\mathrm{break}}})^{-\gamma_1} (E < E_{\mathrm{break}}),\\
        \phi^{\mathrm{Astro}}_{\mathrm{per-flavor}}  \times (\frac{\mathrm{E}}{\mathrm{E}_{\mathrm{break}}})^{-\gamma_2} (E > E_{\mathrm{break}}).
    \end{array}
\right. \space 
\end{split}
\label{eq:bpl}
\end{equation}

This model assumes there are two spectral indexes, one for neutrino energies below an energy break and a second spectral index that extends to higher energies with the normalization defined at the energy break. The parameters to be fit are the flux normalization $\phi^{\mathrm{Astro}}_{\mathrm{per-flavor}}$ ($\nu:\bar\nu$), the energy break $E_{\mathrm{break}}$, and two spectral indices $\gamma_{1}$ and $\gamma_{2}$ with the following best fits:

\begin{equation}
\begin{split}
&\phi^{\mathrm{Astro}}_{\mathrm{per-flavor}} = 1.7 ^{+0.19}_{-0.22}, \quad log_{10}(\frac{\mathrm{E}_\mathrm{break}}{1  \mathrm{GeV}}) \sim 4.36, \\
&\gamma_{1} = 2.79 ^{+0.30}_{-0.50}, \quad \gamma_{2} = 2.52 ^{+0.10}_{-0.09}.
\end{split}
\label{eq:bplresults}
\end{equation}

The BPL model allows a model independent probe of structure in the flux. Structure is expected in some models towards lower energies. For example in some scenarios, the neutrino flux is expected to continue towards lower energies \cite{Murase:2015xka,2011MNRAS.410.2556D} until it reaches an energy break and falls off rapidly to $\gamma \sim 0$ \cite{1990cup..book.....G} below this break.

\begin{figure}[t!]
\includegraphics[width=0.49\textwidth]{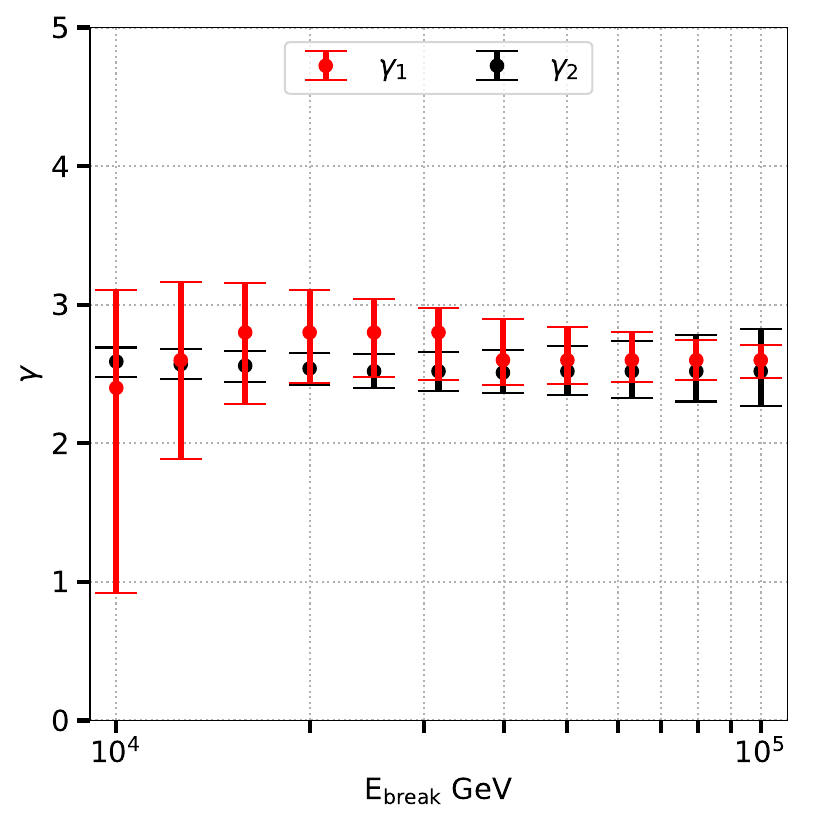}
\caption{Best-fit low energy spectral indexes, $\gamma_1$, calculated assuming various energy breaks. For each point tested, all parameters are allowed to be refit. We lose sensitivity to harder spectrum as the energy break is decreased. We observe the higher energy spectral index to be stable for various models (in black). }
\label{fig:bplscans}
\end{figure}

When fitting a broken power law, we observed a slight softening of the spectrum below the energy break, $\mathrm{E}_\mathrm{break} = 23 \hspace{1mm} \mathrm{TeV}$. The test-statistic that the BPL is preferred over the SPL is 0.4, which is not statistically significant. As a result, $E_{break}$ is poorly constrained, so we quote only the best fit point. The errors on $\gamma_{1}$ and $\gamma_{2}$ are at that fixed $E_{break}$ from one-dimensional profile likelihood scans. We use this data to reject $\gamma_{1} < 0$ to $3.4\sigma$ significance and $\gamma_{1} < 1$ to $3.0\sigma$ at an energy break of 23 TeV. $\gamma_1 < 2$ is only rejected to a $2.1\sigma$ level. A summary of the various energy breaks tested with the corresponding best-fit spectral indexes is shown in Fig. \ref{fig:bplscans}. These results do not indicate any sign of the neutrino flux falling off below 23 TeV. 

An analysis where the BPL model is relevant is that performed by Fang et al. \cite{Fang:2020bkm}. We know the production mechanisms by which neutrinos are produced are from either pp or p$\gamma$ processes. In p$\gamma$ scenarios, protons interact with the photons near the source through photo-pion production only when their energy is above the pion production threshold \cite{Fang:2020bkm}. In these scenarios, Fang et al. use IceCube and Fermi data to predict that the gamma ray flux generated in the source region must cascade down to MeV-GeV energies as to not violate the observed extragalatic Fermi gamma-ray data \cite{Fermi-LAT:2014ryh, Fermi-LAT:2015otn, Fang:2022trf}. The ESTES BPL observations (a lack of hardening in the fitted spectrum below the energy break), under these gamma ray flux interpretations, imply that the dominant neutrino sources are opaque to gamma rays  \cite{Murase:2015xka, Capanema:2020oet, Fang:2022trf}. One example is NGC 1068, an AGN for which IceCube recently reported evidence of TeV neutrino emission that is not matched by a corresponding gamma-ray signal \cite{NGC1068}. 

\subsection{\label{sec:MeasurementHemisphere} Measurement of the diffuse flux assuming a non-isotropic diffuse flux}
We now treat the astrophysical neutrino flux as a sum of two single power laws, one for each hemisphere where the hemisphere is defined using the IceCube local coordinate system. This model is defined as:
\begin{equation}
\begin{split}
&\Phi_{Astro}^{Total}/C_0 = \phi_{Astro,S}  \times (\frac{\mathrm{E}_{\nu}}{100 \mathrm{TeV}})^{-\gamma,S}  (\Theta_\nu < 90^{\circ}) +
\\
& \qquad \qquad \qquad \phi_{Astro,N}  \times (\frac{\mathrm{E}_{\nu}}{100 \mathrm{TeV}})^{-\gamma,N} (\Theta_\nu > 90^{\circ}),
\label{eq:northsouth} 
\\
&\mathrm{where} \hspace{1mm} \mathrm{we} \hspace{1mm} \mathrm{found:}\\
&\phi^{\mathrm{Astro,North}}_{\mathrm{per\ flavor}} = 1.28 ^{+0.83}_{-1.28}, \quad \gamma_{North} = 2.36 ^{+0.24}_{-1.05}, \\
&\phi^{\mathrm{Astro,South}}_{\mathrm{per-flavor}} = 1.56 ^{+0.28}_{-0.24}, \quad \gamma_{South} = 2.66 ^{+0.13}_{-0.16}. \\
\end{split}
\end{equation}

\begin{figure}[t!]
\includegraphics[width=0.49\textwidth]{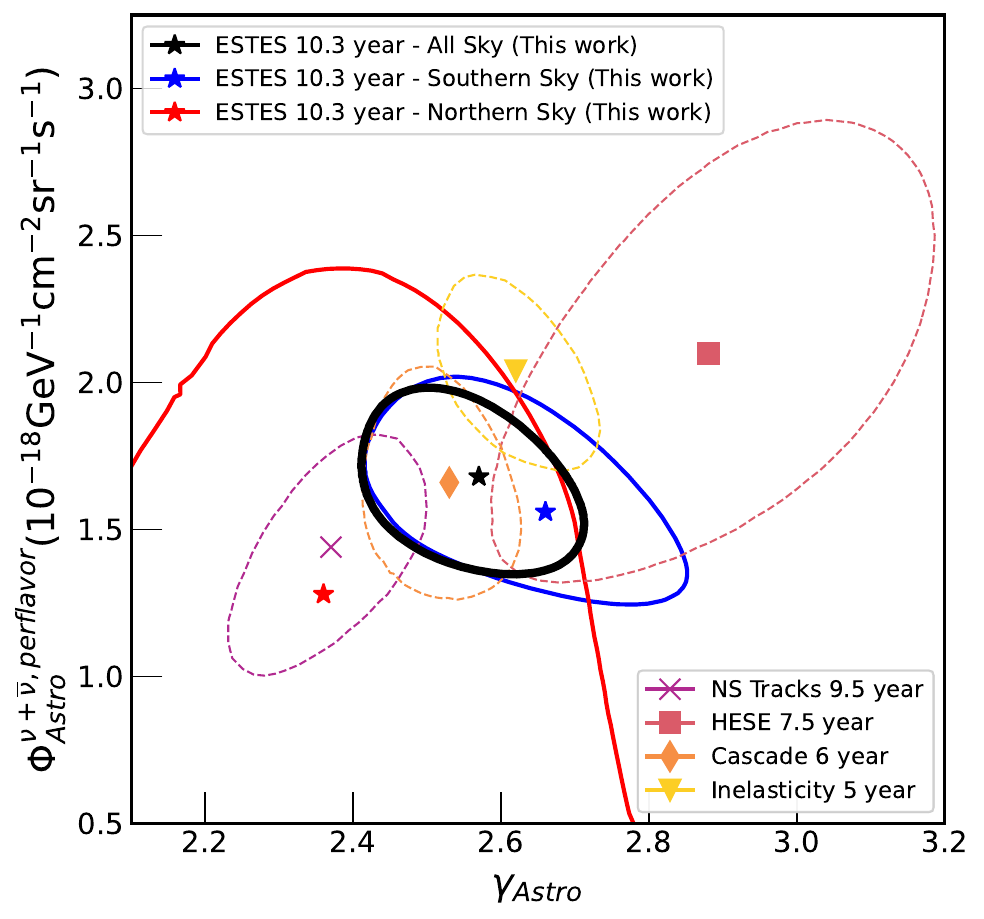}
\caption{A summary of the SPL 68\% confidence intervals for the All-Sky, Southern-Sky, and Northern-Sky SPL measurements. The IceCube ``inelasticity" measurement \cite{inelasticity} is shown as yellow dotted lines. It is the only existing measurement of the flux using starting track events; we note the precision is limited due to harsh cuts in the southern sky. We include recent IceCube results for direct comparison \cite{IceCube:2021uhz,HESENew,SBUCasc}.}
\label{fig:splsummary}
\end{figure}

Given the excellent angular resolution from starting tracks, the hemisphere measurements can be interpreted as independent. The best-fit points and 68\% confidence intervals are shown in Fig. \ref{fig:splsummary} as solid blue and red lines in addition to the isotropic SPL measurement in black.  

We conclude that the fluxes are compatible with each other. It is interesting to see that the southern sky measurement is significantly more constraining - a consequence of the higher proportion of astrophysical neutrinos in the southern sky due to the atmospheric neutrino self-veto.

A smaller, but non-negligible, source of neutrinos from the galactic plane is expected \cite{Gaggero_2015, Albert_2018, 2017ApJ...849...67A, Aartsen_2019} below 100 TeV. The Fermi-LAT benchmark model \cite{fermitlat,fermipi0} describes the diffuse emission of gamma rays likely due to the interaction of cosmic rays with the interstellar medium (ISM) (or surrounding sources). In these interactions, both charged and neutral pions are produced. The neutral pions $(\pi^{0})$ decay into a photon pair while the charged pions $(\pi^{\pm})$ decay into neutrinos and muons. Therefore, the expected diffuse neutrino flux from the galactic plane is closely connected to gamma ray measurements. This neutrino model is referred to as the Fermi-$\pi^{0}$ model as described in reference \cite{IceCube:2017trr}. The spatial distribution of the gamma rays is considered and it is further combined with a neutrino single power law flux of E$^{-2.7}$.

\begin{figure}[t!]
\includegraphics[width=0.45\textwidth]{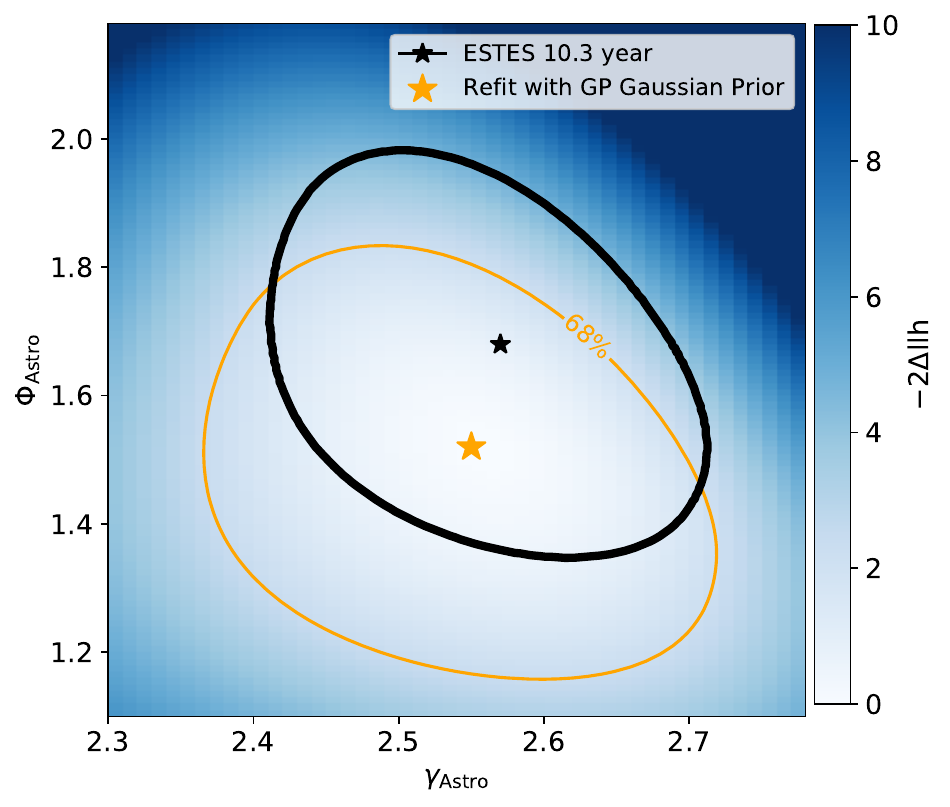}
\caption{Single power law flux measurement as described in Sec. \ref{sec:Measurement} (in black) and introducing the galactic plane as additional nuisance parameter in the fit.}
\label{fig:fermipi0}
\end{figure}

We directly test the impact of the galactic plane on the isotropic diffuse flux measurement, treating the Fermi-$\pi^0$ flux normalization as a Gaussian nuisance parameter using the measured fluxes from an IceCube dedicated search for neutrinos from the galactic plane \cite{dnncascade}. Figure \ref{fig:fermipi0} shows the measured isotropic diffuse flux after including the Fermi-$\pi^{0}$ term and fitting to the data again. We observe at most a $10\%$ impact on the isotropic normalization with negligible impact on the spectral index. The same treatment was performed for the segmented flux measurement. The flux normalization shifts are shown in Tab. \ref{tab:segmentedresults}. This treatment was also repeated for the broken power law measurement \ref{sec:MeasurementBPL} and hemisphere measurement\ref{sec:MeasurementHemisphere} with negligible impact. Given the large errors introduced by adding the galactic plane diffuse model, an extension of this measurement including right ascension is motivated but beyond the scope of this work. 

\subsection{\label{sec:Prompt} Search for prompt atmospheric neutrinos }
Atmospheric neutrinos resulting from the decay of charmed mesons in atmospheric showers are referred to as prompt neutrinos. The charmed mesons decay promptly resulting in a harder spectrum than their conventional counterparts. We treat the prompt neutrino flux as an independent parameter with the cosmic ray model and self-veto uncertainties applied as described in Sec. \ref{sec:SystematicUncertainties}. The assumed cosmic-ray flux model is Gaisser H4a-GST with Sibyll 2.3c. For reference, the theoretical BERSS flux \cite{Bhattacharya:2016jce} is $\sim3\times$ smaller at 50 TeV and the ERS flux \cite{Enberg:2008te} is similar to the theoretical prompt flux tested here. We do not observe any evidence for the prompt flux and show the test-statistic in Tab. \ref{tab:promptlimits}. We searched for the prompt flux assuming a single power law and broken power law flux hypothesis from Sec. \ref{sec:measurement} and Sec. \ref{sec:MeasurementBPL}, respectively setting limits on the prompt flux under both of these astrophysical flux models. The limits shown correspond to a scaling factor multiplied by the prompt flux shown in Fig. \ref{fig:CRHImodels}. 

\renewcommand{\arraystretch}{1.5} 
\begin{table}[h]
\begin{tabular}{ @{\hspace{15pt}}r @{\hspace{15pt}}| @{\hspace{15pt}}c @{\hspace{15pt}}l @{\hspace{15pt}}| @{\hspace{15pt}}c @{\hspace{15pt}}l } 
\hline
\hline
UL $\times \Phi_\mathrm{prompt}$ & 90\% Upper Limit \\
\hline 
Single Power Law & 3.19 \\ 
Broken Power Law & 3.20 \\ 
\hline
\hline
\end{tabular}
\caption{Prompt upper limits as computed under the single power law and broken power law flux astrophysical models. We observe similar results with a slightly worse limit for the BPL due to the additional flexibility in the model.}
\label{tab:promptlimits}
\end{table}

\subsection{\label{sec:IceCubeSummary} Diffuse flux measurement summary and outlook}
Figure \ref{fig:splsummary} shows a summary of all recent IceCube measurements of the astrophysical diffuse flux. Most importantly, it shows that despite numerous techniques employed over the past decade, the different neutrino data sets converge towards similar results. 

The HESE 7.5 year measurement \cite{HESENew} focuses on high energy starting events. The dataset is dominated by cascade-like events with a non-negligible contribution from starting track-like events (17\% starting tracks). The HESE analysis applies a minimum energy cut at 60 TeV limiting the measurement of the astrophysical flux to higher energy. This also limits the statistics and correspondingly leads to a measurement that is statistically limited (largest contour in Fig. \ref{fig:splsummary}). 

The Cascade 6-year and 5-year measurements \cite{SBUCasc, inelasticity} are driven by cascade events extending IceCube's sensitivity to the astrophysical flux down to 16 TeV. These lower energy measurements take advantage of the self-veto effect, assuming a muon response modeled as a fixed step-function. While these measurements are greatly constraining, we now believe that any measurement utilizing the self-veto effect should be inclusive of uncertainties from the choice of self-veto flux model.  

The Northern Sky Tracks 9.5-year measurement uses through going and starting muon tracks in the northern equatorial sky ($\theta > 85^{\circ}$). 
This measurement is limited by the energy resolution of the through-going events because the reconstructed muon energy can only be interpreted as a lower limit on the expected neutrino energy. The zenith angle cut also means there is no self-veto effect to take advantage of. However, the high statistics and wide energy range of this well studied event selection makes it a powerful measurement.

The ESTES 10.3-year measurement (this work) searches for starting track-like events over the entire sky for energies above 1 TeV. This event selection takes advantage of the excellent energy and directional resolution of such an event morphology. In the southern sky, the self-veto effect improves the astrophysical neutrino purity of the selection.

\section{\label{sec:Conclusion} Conclusion}
A measurement of the astrophysical diffuse neutrino flux was presented in this work using novel techniques. This paper outlined the construction of the ESTES dataset, its performance, and the measurement of the diffuse flux from the ESTES selection applied to 10.3 years of IceCube data. We also outlined the new systematic uncertainty terms: self-veto effect and hadronic interaction model gradient which were shown to contribute non-negligible impacts on the astrophysical flux measurement. A search for the atmospheric prompt neutrino flux was also presented. No evidence for the prompt flux was found and upper limits were set on the Gaisser H4a-GST cosmic ray model with Sibyll 2.3c prompt flux model of 3.2 times the theoretical prediction. The amplitude of the prompt flux remains one of the unresolved mysteries in the diffuse neutrino sky.

The ESTES dataset of 10,798 events was extracted using veto-techniques and boosted decision trees to search for starting track events in the northern and southern hemisphere. The overwhelming atmospheric muon background was successfully reduced from 3\,kHz down to 1\,$\mu$Hz ($<1\%$ of the remaining data) while retaining a large effective area for neutrinos in the southern hemisphere.  The dataset observed at least 10,000 neutrino events of which 1000 were localized to the southern sky. This dataset opens up the possibility of conducting other neutrino studies, such as searching for neutrino sources in the southern sky.

The ESTES dataset was used to search for, and characterize, the astrophysical neutrino flux. Both a single power law and broken power law form for the astrophysical flux were fitted. The best-fit spectral index for the single power law fit is $\gamma = 2.58 ^{+0.10}_{-0.09}$ and per-flavor normalization is $\Phi^{\mathrm{Astro}}_{\mathrm{per\ flavor}} = 1.68 ^{+0.19}_{-0.22}$ (at 100\,TeV). The sensitive energy range for this particular flux model is 3-550\,TeV, marking the first time the neutrino flux is measured to such precision below 16 TeV. The observation of the diffuse flux below 100 TeV is in agreement with, and independent from, IceCube's 6-year cascade-event based result \cite{SBUCasc}. Assuming the single power law flux, we then presented a segmented measurement of the normalization from 300 GeV to 100 PeV showing consistent normalization with the measured single power law.

We tested the impact of the galactic plane under the Fermi $\pi^0$ flux model and concluded that while the expected impact on the diffuse flux spectral index is at the sub-percent level it can still contribute to the overall normalization by $\sim 10\%$.

Finally, a measurement of the flux under the broken power law assumption was performed. We tested a lower (higher) energy spectral index below (above) a break energy. We are able to reject $\gamma_1 < 1$ to greater than $3\sigma$ significance and $\gamma_1 < 2$ to $2.1\sigma$ significance for energies below 23 TeV. At 40 TeV, we measure $\gamma_1 = 2.6 ^{+0.3}_{-0.2}$ which is largely consistent with IceCube's 6-year result using cascades \cite{SBUCasc}. Overall, we do not observe a departure from a single power law at lower energies.

In conclusion, we present a measurement of the diffuse flux over the entire sky from 300 GeV to 100 PeV using the starting track event morphology. This is the first measurement of starting tracks below 100 TeV, allowing us to study the diffuse astrophysical neutrino flux with a new sample at lower energies. No evidence was found for structure in the flux beyond a single power law spanning from 3 TeV to 550 TeV.

\begin{acknowledgments}
The IceCube collaboration acknowledges the significant contributions to this manuscript from Sarah Mancina, Jesse Osborn, and Manuel Silva. 
\\
The authors gratefully acknowledge the support from the following agencies and institutions: USA {\textendash} U.S. National Science Foundation-Office of Polar Programs,
U.S. National Science Foundation-Physics Division,
U.S. National Science Foundation-EPSCoR,
U.S. National Science Foundation-Office of Advanced Cyberinfrastructure,
Wisconsin Alumni Research Foundation,
Center for High Throughput Computing (CHTC) at the University of Wisconsin{\textendash}Madison,
Open Science Grid (OSG),
Partnership to Advance Throughput Computing (PATh),
Advanced Cyberinfrastructure Coordination Ecosystem: Services {\&} Support (ACCESS),
Frontera computing project at the Texas Advanced Computing Center,
U.S. Department of Energy-National Energy Research Scientific Computing Center,
Particle astrophysics research computing center at the University of Maryland,
Institute for Cyber-Enabled Research at Michigan State University,
Astroparticle physics computational facility at Marquette University,
NVIDIA Corporation,
and Google Cloud Platform;
Belgium {\textendash} Funds for Scientific Research (FRS-FNRS and FWO),
FWO Odysseus and Big Science programmes,
and Belgian Federal Science Policy Office (Belspo);
Germany {\textendash} Bundesministerium f{\"u}r Bildung und Forschung (BMBF),
Deutsche Forschungsgemeinschaft (DFG),
Helmholtz Alliance for Astroparticle Physics (HAP),
Initiative and Networking Fund of the Helmholtz Association,
Deutsches Elektronen Synchrotron (DESY),
and High Performance Computing cluster of the RWTH Aachen;
Sweden {\textendash} Swedish Research Council,
Swedish Polar Research Secretariat,
Swedish National Infrastructure for Computing (SNIC),
and Knut and Alice Wallenberg Foundation;
European Union {\textendash} EGI Advanced Computing for research;
Australia {\textendash} Australian Research Council;
Canada {\textendash} Natural Sciences and Engineering Research Council of Canada,
Calcul Qu{\'e}bec, Compute Ontario, Canada Foundation for Innovation, WestGrid, and Digital Research Alliance of Canada;
Denmark {\textendash} Villum Fonden, Carlsberg Foundation, and European Commission;
New Zealand {\textendash} Marsden Fund;
Japan {\textendash} Japan Society for Promotion of Science (JSPS)
and Institute for Global Prominent Research (IGPR) of Chiba University;
Korea {\textendash} National Research Foundation of Korea (NRF);
Switzerland {\textendash} Swiss National Science Foundation (SNSF).
\end{acknowledgments}

\appendix

\section{\label{sec:SegmentedAppendix} Segmented Power Law Validation}
\begin{figure*}[p!]
    \centering
    \begin{minipage}[b]{.49\textwidth}
        \centering
        \includegraphics[width=.95\textwidth]{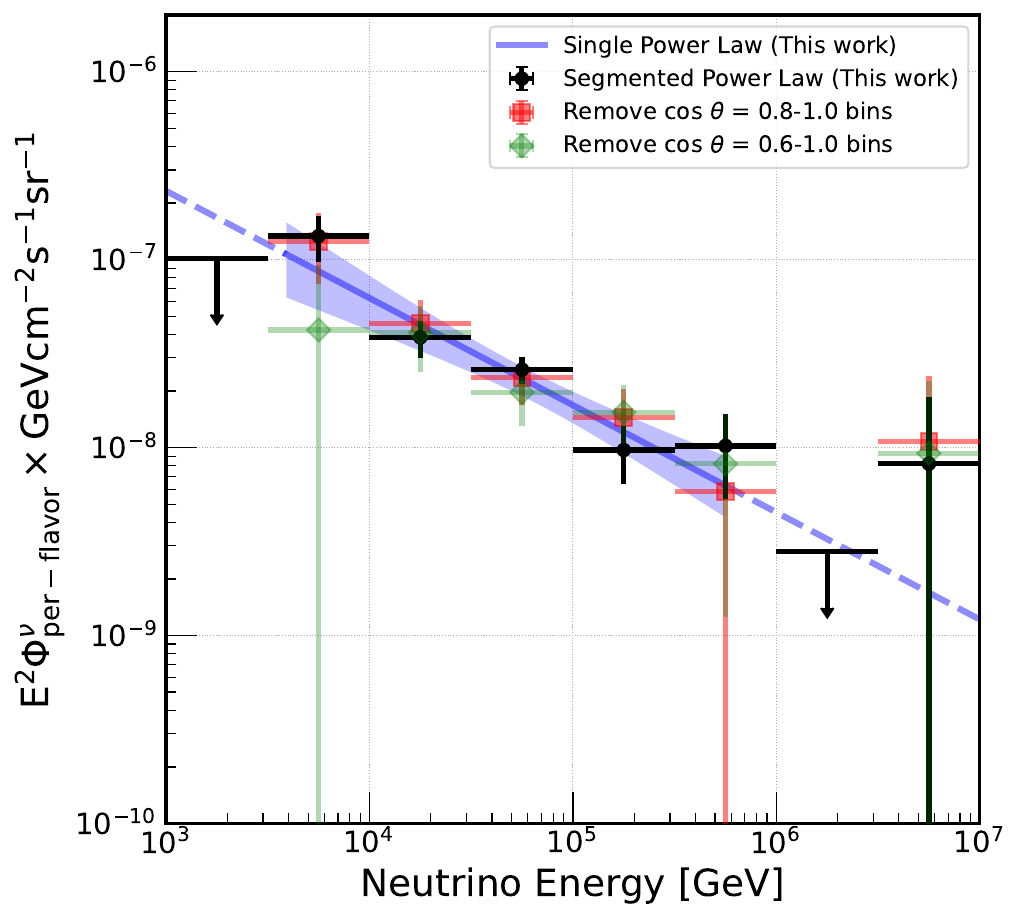}
        \caption{Distribution of the unfolded flux as first shown in Fig. \ref{fig:fluxsummary} and the unfolded fluxes after removing the most vertical bins from the measurement. This cross-check was performed to test the isotropic-qualities of all segments. We observe a stable flux over all energies.}
        \label{fig:sanitycheckunfolded}
    \end{minipage}
    \hspace{1mm}
    \begin{minipage}[b]{.49\textwidth}
        \centering
        \includegraphics[width=.95\textwidth]{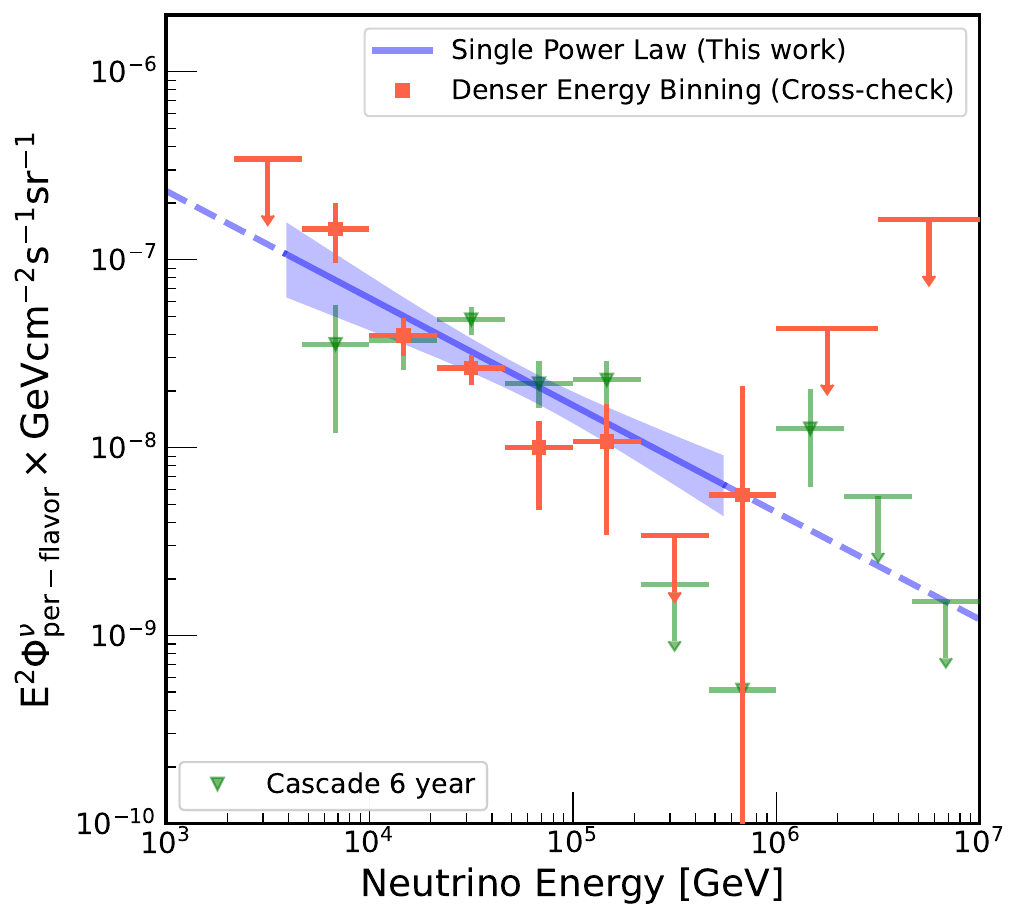}
        \caption{Distribution of the unfolded flux as first shown in Fig. \ref{fig:fluxsummary}. The red points are the segmented flux measurement using denser energy binning as a cross-check to compare directly with 6-year IceCube cascade result \cite{SBUCasc}. There is no sensitivity to the astrophysical flux in the 1-2.15 TeV bin.}
        \label{fig:segmented_dense_binning_cross_check}
    \end{minipage}
\end{figure*}

Figure \ref{fig:sanitycheckunfolded} shows a check for the stability of each segment over two portions of the sky. In this test, the most vertical bins are removed and the segmented flux is recomputed for the same segments. We note all segments remain consistent despite the reduction in sample size. However, we also note that the 3-10 TeV segment decreased by more than 1$\sigma$. We found the most dominant background to astrophysical neutrinos at such energies and zeniths to be from mis-reconstructed atmospheric neutrinos from the horizon. A full set of figures is found in App. \ref{sec:TrueNeutrinoZeniths}. Future iterations of this analysis should be performed using more robust directional reconstructions to reduce this background. 

To directly compare to the segmented power law measurement from the 6-year IceCube cascade result\cite{SBUCasc}, the segmented fit was performed again using the same energy bins. The result is shown in Fig. \ref{fig:segmented_dense_binning_cross_check}. The tension in the 4.64-10 TeV bin is 2.3$\sigma$. The maximum tension per bin is in the 21.5-46.4 TeV bin at 3.7$\sigma$ (omitting trials correction factor). We observe a ``dip"-like structure between 215-464 TeV (similar to that observed with cascades), but note that the reported upper limit is worse despite the improved livetime of this dataset. The larger number of bins leads to increased correlations of up to $\pm$25-30\% between bins.

\section{\label{sec:DetectorSystematicPriors} Detector Systematic Constraints}
\begin{figure*}[b]
\includegraphics[width=0.45\textwidth]{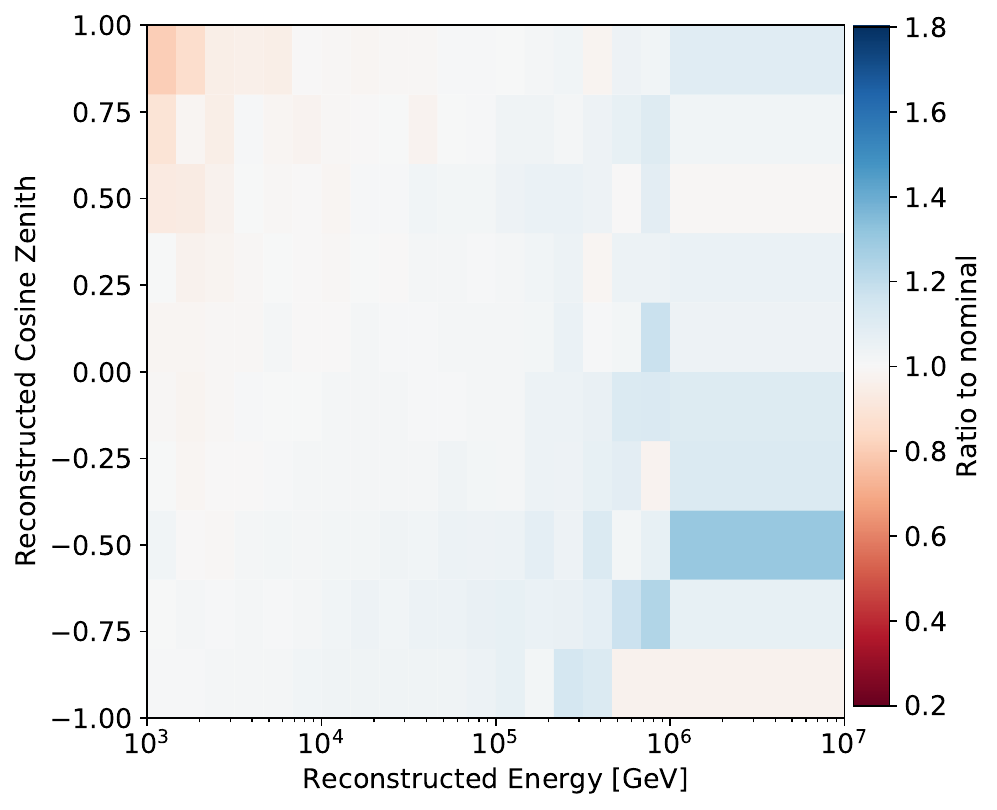}
\includegraphics[width=0.45\textwidth]{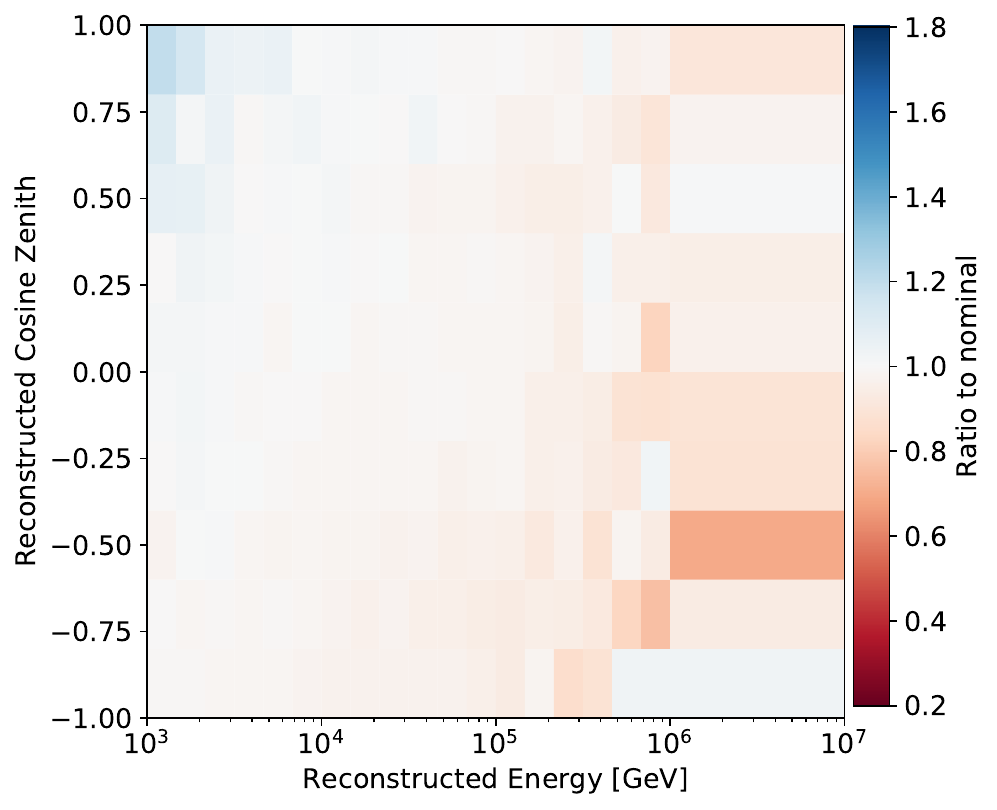}
\caption{Scattering coefficients shifted by +1 and -1 $\sigma$.}
\label{fig:scat}
\end{figure*}

\begin{figure*}[b]
\includegraphics[width=0.45\textwidth]{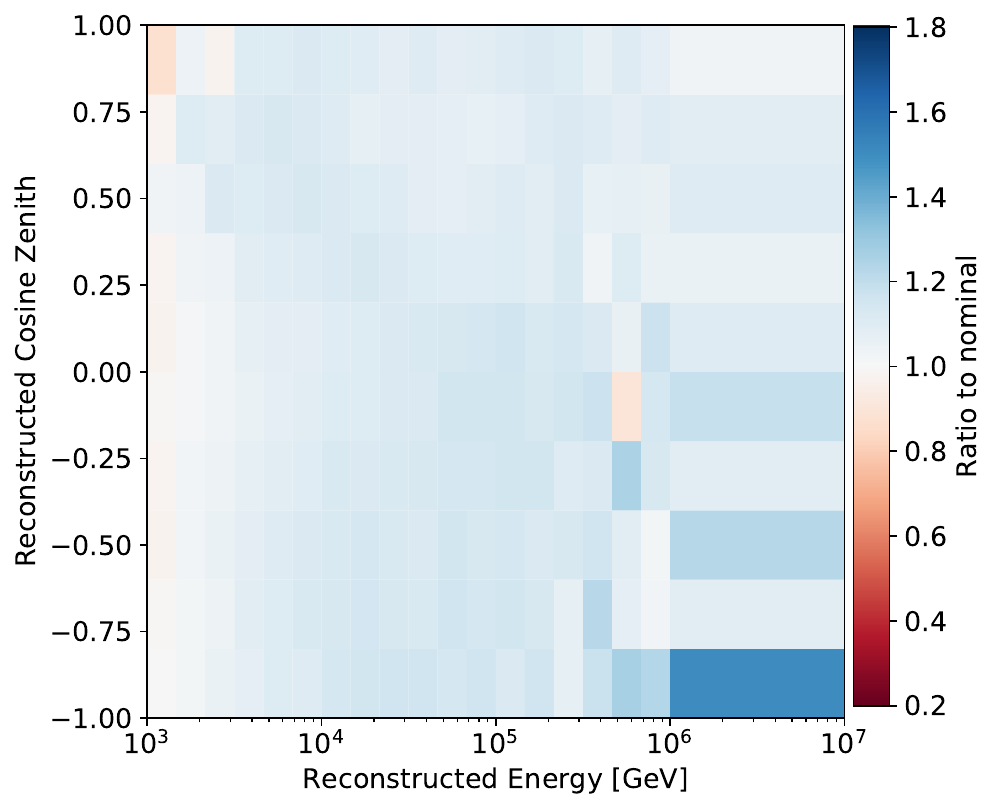}
\includegraphics[width=0.45\textwidth]{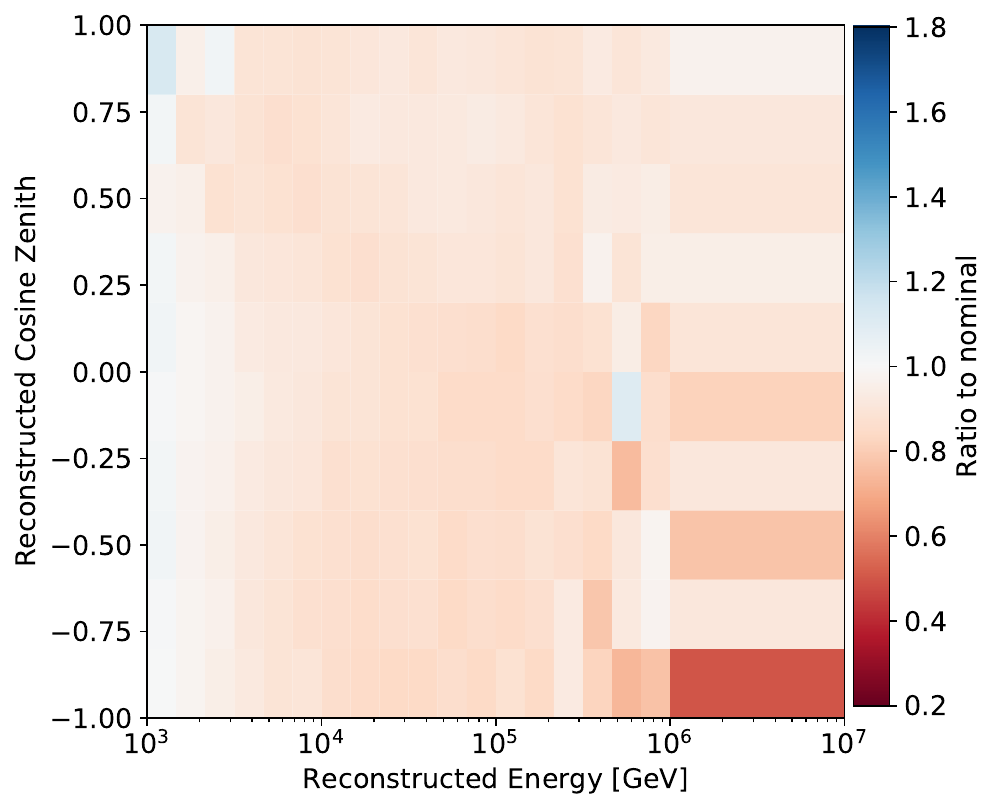}
\caption{Absorption coefficients shifted by +1 and -1 $\sigma$.}
\label{fig:abs}
\end{figure*}

\begin{figure*}[b]
\includegraphics[width=0.45\textwidth]{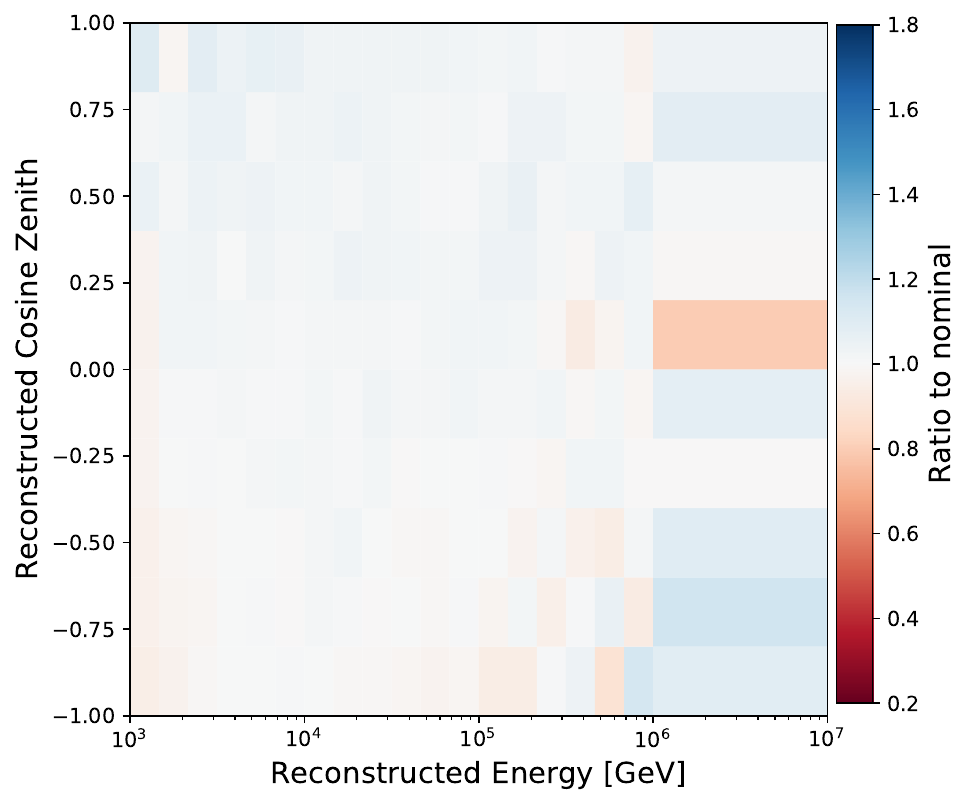}
\includegraphics[width=0.45\textwidth]{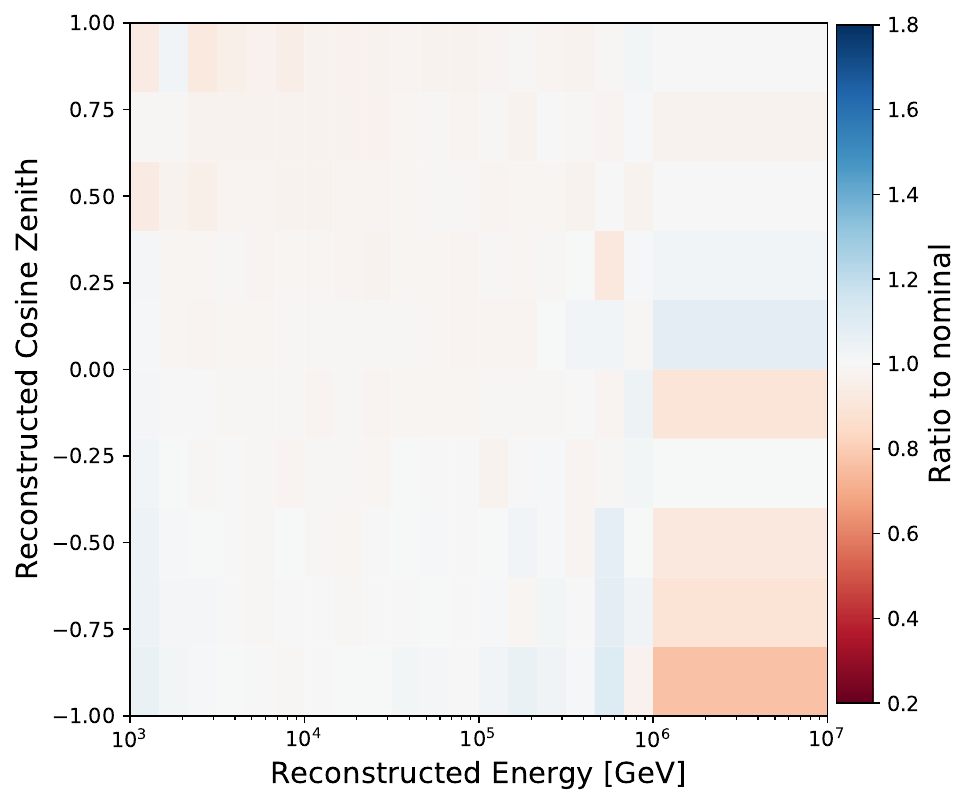}
\caption{Hole-ice p$_{0}$ parameter shifted by +1 and -1 $\sigma$.}
\label{fig:p0}
\end{figure*}

\begin{figure*}[b]
\includegraphics[width=0.45\textwidth]{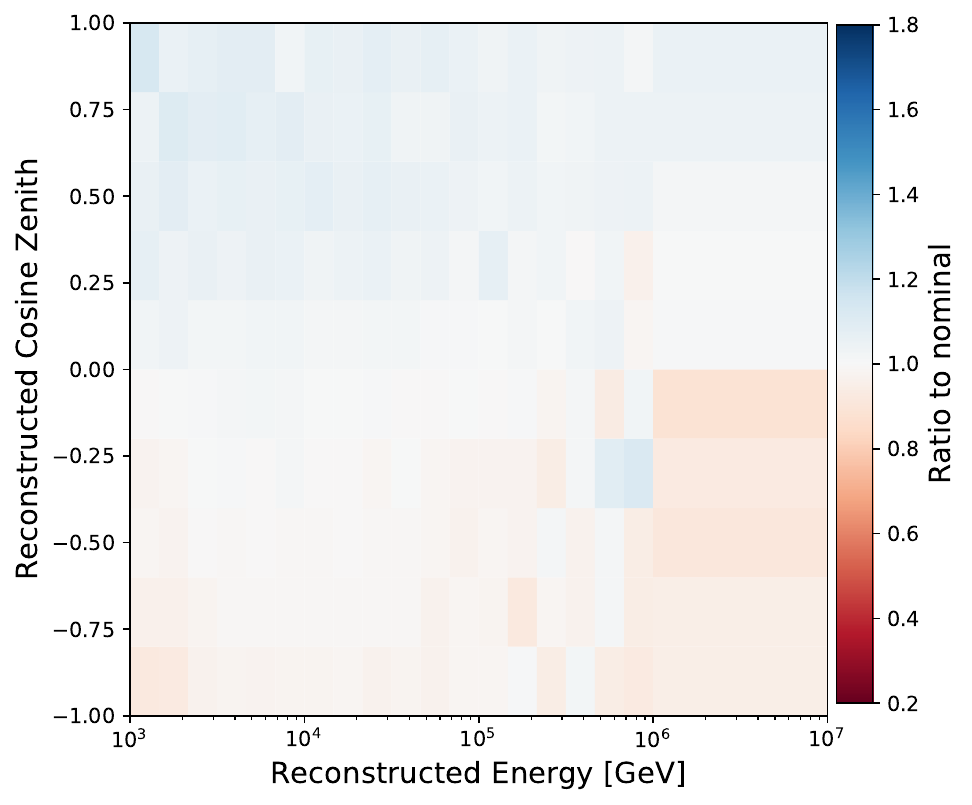}
\includegraphics[width=0.45\textwidth]{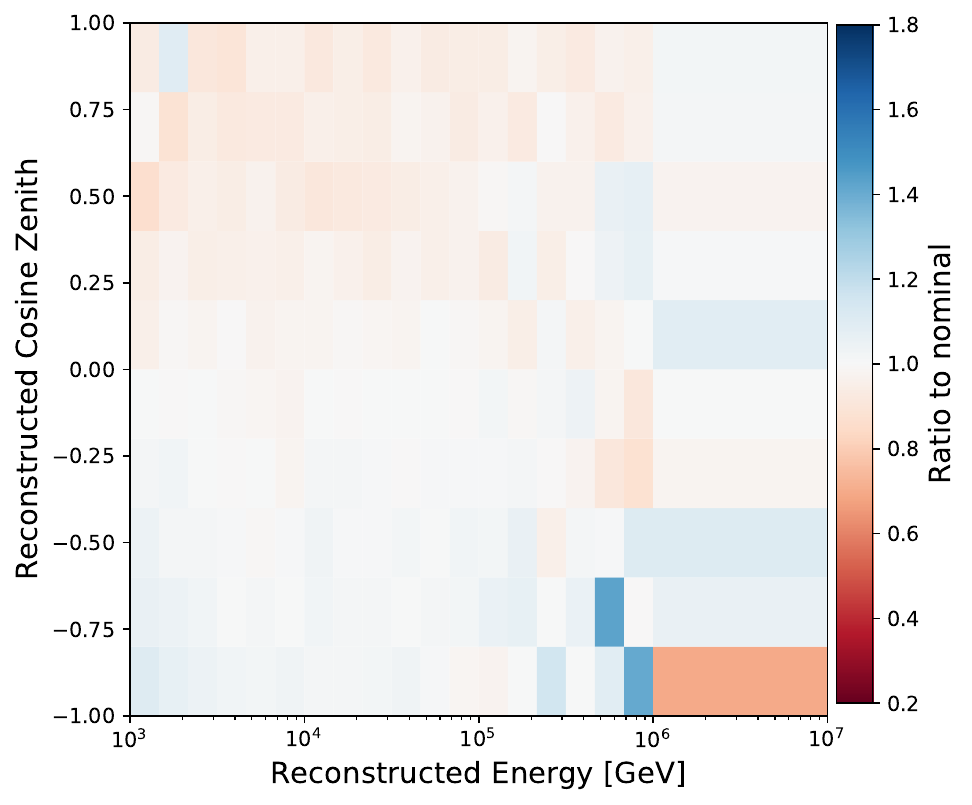}
\caption{Hole-ice p$_{1}$ parameter shifted by +1 and -1 $\sigma$.}
\label{fig:p1}
\end{figure*}

\begin{figure*}[b]
\includegraphics[width=0.45\textwidth]{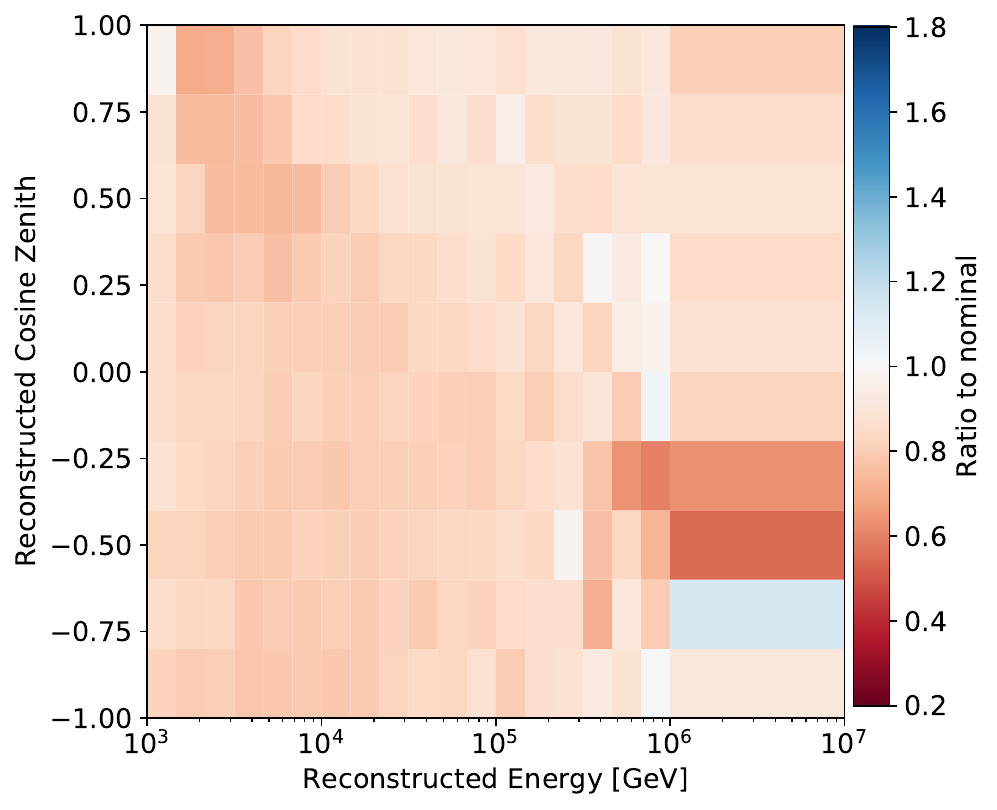}
\includegraphics[width=0.45\textwidth]{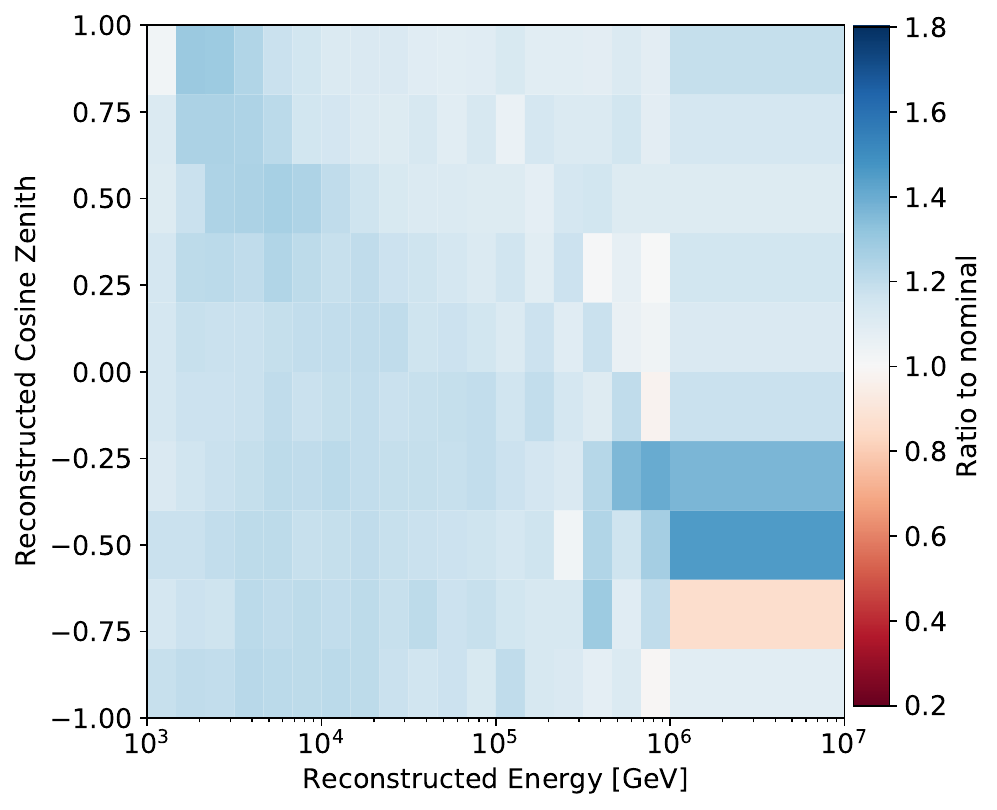}
\caption{DOM overall efficiency shifted by +1 and -1 $\sigma$.}
\label{fig:domeff}
\end{figure*}

The detector systematics are treated as uncorrelated parameters in the likelihood. When applicable, a Gaussian penalty term is added to the likelihood to better inform our model of external measurements. The same set of neutrino events as described in Section \ref{sec:Overview} are propagated through the event selection as described in Section \ref{sec:EventSelection}. The energy and directional reconstructions are then reevaluated under these new detector configurations. The changes in the detector response are shown as $\pm 1 \sigma$ shifts in the event expectation with respect to their nominal expectation in Fig. \ref{fig:scat} through Fig. \ref{fig:domeff}. The 2D MC templates are only shown at their $\pm1\sigma$ point but intermediate points were also simulated such that we construct a 1D parametrization of the detector response per systematic uncertainty.

\section{\label{sec:TrueNeutrinoZeniths} Predicted Neutrino Zeniths}

\begin{figure*}[t]
\centering
\begin{tabular}{cc}
\includegraphics[width=0.33\linewidth]{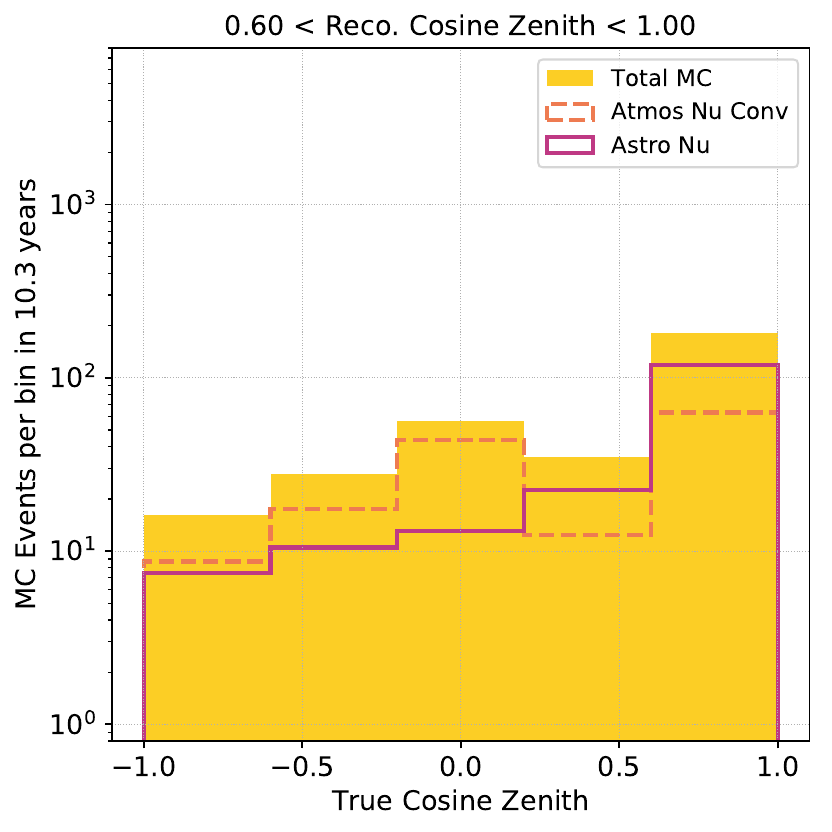} &   \includegraphics[width=0.33\linewidth]{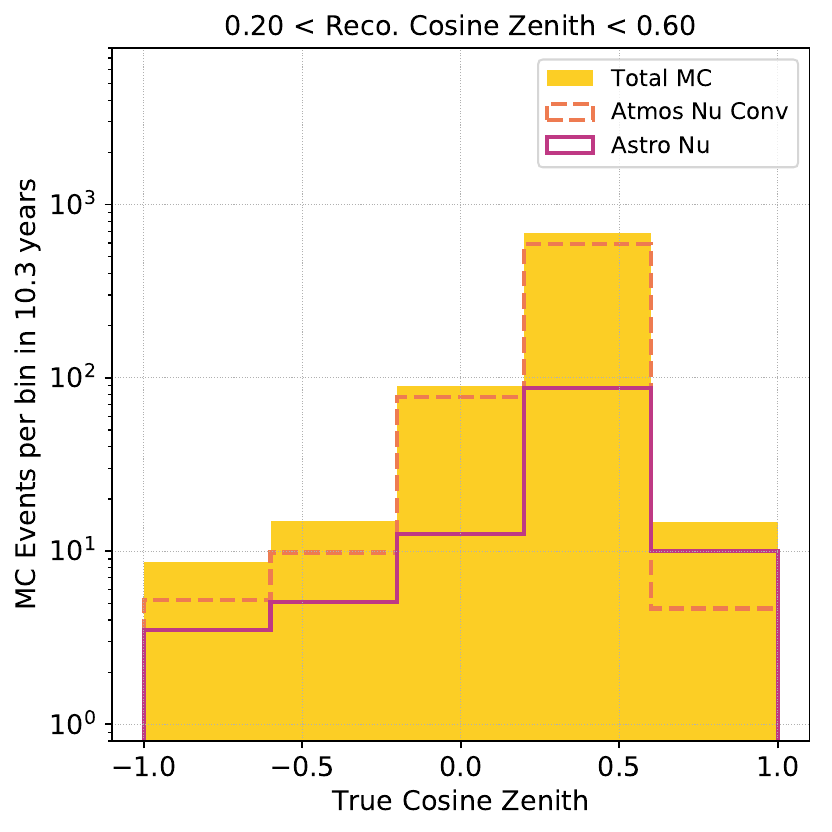} \\
\includegraphics[width=0.33\linewidth]{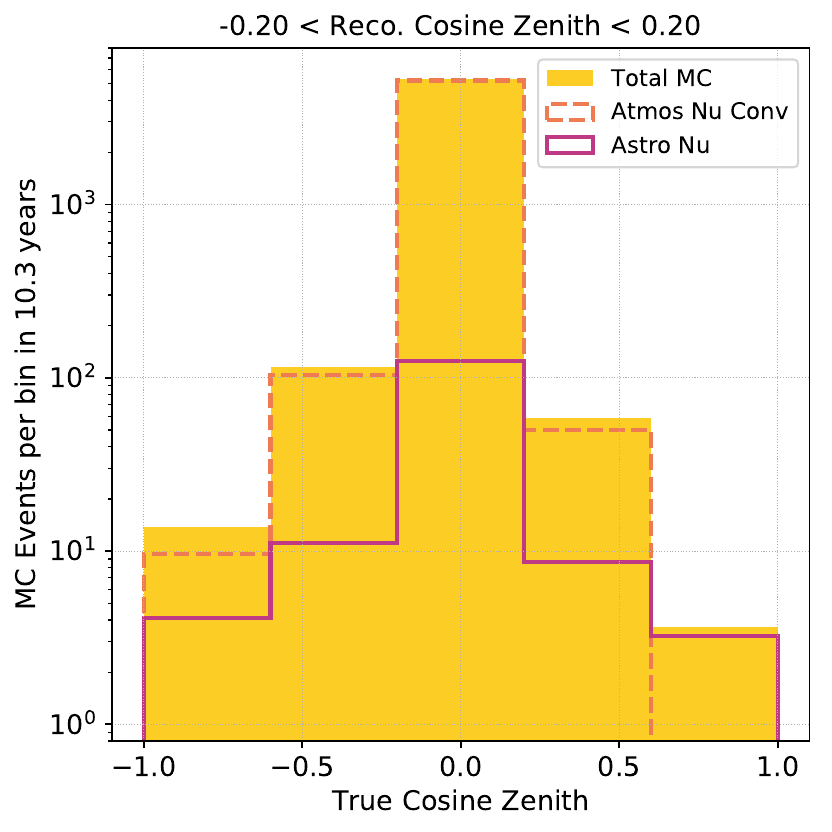} &
\includegraphics[width=0.33\linewidth]{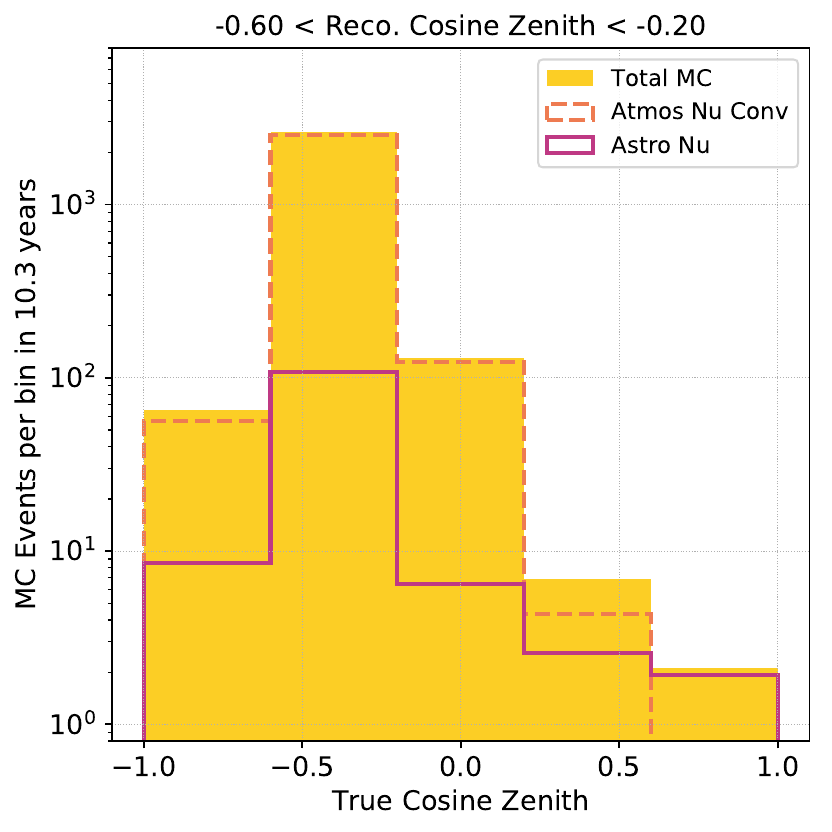} \\
\multicolumn{2}{c}{
\includegraphics[width=0.33\linewidth]{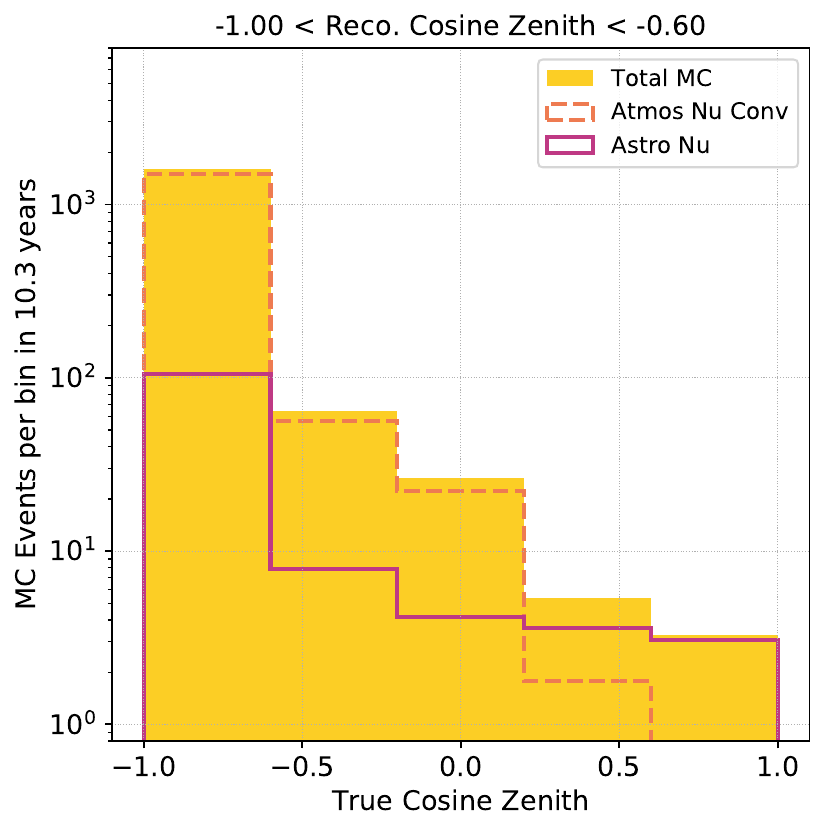} 
} \\
\end{tabular}
\caption{True cosine zenith distributions using the neutrino simulation. Each sub-figure is binned by reconstructed cosine zenith distribution. We draw attention here to the dominant background in reconstructed zenith = 0.6 - 1.0. These events are determined to be a mixture of true neutral current events and high inelasticity muon neutrinos in the charged current channel. Future iterations of this dataset can improve purity by applying more robust directional reconstructions.}
\label{fig:trueneutrinozeniths}
\end{figure*}

The true cosine zenith distributions are shown in Fig. \ref{fig:trueneutrinozeniths}. Cuts for each subplot are made on the reconstructed zenith used as an observable in the measurement. The simulated rates take into account the parameters from Tab. \ref{tab:systs}.

\twocolumngrid
\nocite{*}
\bibliographystyle{apsrev4-2}
\bibliography{main_ESTES_overview}

\end{document}

%% file: authorlist.tex
\affiliation{III. Physikalisches Institut, RWTH Aachen University, D-52056 Aachen, Germany}
\affiliation{Department of Physics, University of Adelaide, Adelaide, 5005, Australia}
\affiliation{Dept. of Physics and Astronomy, University of Alaska Anchorage, 3211 Providence Dr., Anchorage, AK 99508, USA}
\affiliation{Dept. of Physics, University of Texas at Arlington, 502 Yates St., Science Hall Rm 108, Box 19059, Arlington, TX 76019, USA}
\affiliation{CTSPS, Clark-Atlanta University, Atlanta, GA 30314, USA}
\affiliation{School of Physics and Center for Relativistic Astrophysics, Georgia Institute of Technology, Atlanta, GA 30332, USA}
\affiliation{Dept. of Physics, Southern University, Baton Rouge, LA 70813, USA}
\affiliation{Dept. of Physics, University of California, Berkeley, CA 94720, USA}
\affiliation{Lawrence Berkeley National Laboratory, Berkeley, CA 94720, USA}
\affiliation{Institut f{\"u}r Physik, Humboldt-Universit{\"a}t zu Berlin, D-12489 Berlin, Germany}
\affiliation{Fakult{\"a}t f{\"u}r Physik {\&} Astronomie, Ruhr-Universit{\"a}t Bochum, D-44780 Bochum, Germany}
\affiliation{Universit{\'e} Libre de Bruxelles, Science Faculty CP230, B-1050 Brussels, Belgium}
\affiliation{Vrije Universiteit Brussel (VUB), Dienst ELEM, B-1050 Brussels, Belgium}
\affiliation{Department of Physics and Laboratory for Particle Physics and Cosmology, Harvard University, Cambridge, MA 02138, USA}
\affiliation{Dept. of Physics, Massachusetts Institute of Technology, Cambridge, MA 02139, USA}
\affiliation{Dept. of Physics and The International Center for Hadron Astrophysics, Chiba University, Chiba 263-8522, Japan}
\affiliation{Department of Physics, Loyola University Chicago, Chicago, IL 60660, USA}
\affiliation{Dept. of Physics and Astronomy, University of Canterbury, Private Bag 4800, Christchurch, New Zealand}
\affiliation{Dept. of Physics, University of Maryland, College Park, MD 20742, USA}
\affiliation{Dept. of Astronomy, Ohio State University, Columbus, OH 43210, USA}
\affiliation{Dept. of Physics and Center for Cosmology and Astro-Particle Physics, Ohio State University, Columbus, OH 43210, USA}
\affiliation{Niels Bohr Institute, University of Copenhagen, DK-2100 Copenhagen, Denmark}
\affiliation{Dept. of Physics, TU Dortmund University, D-44221 Dortmund, Germany}
\affiliation{Dept. of Physics and Astronomy, Michigan State University, East Lansing, MI 48824, USA}
\affiliation{Dept. of Physics, University of Alberta, Edmonton, Alberta, T6G 2E1, Canada}
\affiliation{Erlangen Centre for Astroparticle Physics, Friedrich-Alexander-Universit{\"a}t Erlangen-N{\"u}rnberg, D-91058 Erlangen, Germany}
\affiliation{Physik-department, Technische Universit{\"a}t M{\"u}nchen, D-85748 Garching, Germany}
\affiliation{D{\'e}partement de physique nucl{\'e}aire et corpusculaire, Universit{\'e} de Gen{\`e}ve, CH-1211 Gen{\`e}ve, Switzerland}
\affiliation{Dept. of Physics and Astronomy, University of Gent, B-9000 Gent, Belgium}
\affiliation{Dept. of Physics and Astronomy, University of California, Irvine, CA 92697, USA}
\affiliation{Karlsruhe Institute of Technology, Institute for Astroparticle Physics, D-76021 Karlsruhe, Germany}
\affiliation{Karlsruhe Institute of Technology, Institute of Experimental Particle Physics, D-76021 Karlsruhe, Germany}
\affiliation{Dept. of Physics, Engineering Physics, and Astronomy, Queen's University, Kingston, ON K7L 3N6, Canada}
\affiliation{Department of Physics {\&} Astronomy, University of Nevada, Las Vegas, NV 89154, USA}
\affiliation{Nevada Center for Astrophysics, University of Nevada, Las Vegas, NV 89154, USA}
\affiliation{Dept. of Physics and Astronomy, University of Kansas, Lawrence, KS 66045, USA}
\affiliation{Centre for Cosmology, Particle Physics and Phenomenology - CP3, Universit{\'e} catholique de Louvain, Louvain-la-Neuve, Belgium}
\affiliation{Department of Physics, Mercer University, Macon, GA 31207-0001, USA}
\affiliation{Dept. of Astronomy, University of Wisconsin{\textemdash}Madison, Madison, WI 53706, USA}
\affiliation{Dept. of Physics and Wisconsin IceCube Particle Astrophysics Center, University of Wisconsin{\textemdash}Madison, Madison, WI 53706, USA}
\affiliation{Institute of Physics, University of Mainz, Staudinger Weg 7, D-55099 Mainz, Germany}
\affiliation{Department of Physics, Marquette University, Milwaukee, WI 53201, USA}
\affiliation{Institut f{\"u}r Kernphysik, Westf{\"a}lische Wilhelms-Universit{\"a}t M{\"u}nster, D-48149 M{\"u}nster, Germany}
\affiliation{Bartol Research Institute and Dept. of Physics and Astronomy, University of Delaware, Newark, DE 19716, USA}
\affiliation{Dept. of Physics, Yale University, New Haven, CT 06520, USA}
\affiliation{Columbia Astrophysics and Nevis Laboratories, Columbia University, New York, NY 10027, USA}
\affiliation{Dept. of Physics, University of Oxford, Parks Road, Oxford OX1 3PU, United Kingdom}
\affiliation{Dipartimento di Fisica e Astronomia Galileo Galilei, Universit{\`a} Degli Studi di Padova, I-35122 Padova PD, Italy}
\affiliation{Dept. of Physics, Drexel University, 3141 Chestnut Street, Philadelphia, PA 19104, USA}
\affiliation{Physics Department, South Dakota School of Mines and Technology, Rapid City, SD 57701, USA}
\affiliation{Dept. of Physics, University of Wisconsin, River Falls, WI 54022, USA}
\affiliation{Dept. of Physics and Astronomy, University of Rochester, Rochester, NY 14627, USA}
\affiliation{Department of Physics and Astronomy, University of Utah, Salt Lake City, UT 84112, USA}
\affiliation{Dept. of Physics, Chung-Ang University, Seoul 06974, Republic of Korea}
\affiliation{Oskar Klein Centre and Dept. of Physics, Stockholm University, SE-10691 Stockholm, Sweden}
\affiliation{Dept. of Physics and Astronomy, Stony Brook University, Stony Brook, NY 11794-3800, USA}
\affiliation{Dept. of Physics, Sungkyunkwan University, Suwon 16419, Republic of Korea}
\affiliation{Institute of Physics, Academia Sinica, Taipei, 11529, Taiwan}
\affiliation{Dept. of Physics and Astronomy, University of Alabama, Tuscaloosa, AL 35487, USA}
\affiliation{Dept. of Astronomy and Astrophysics, Pennsylvania State University, University Park, PA 16802, USA}
\affiliation{Dept. of Physics, Pennsylvania State University, University Park, PA 16802, USA}
\affiliation{Dept. of Physics and Astronomy, Uppsala University, Box 516, SE-75120 Uppsala, Sweden}
\affiliation{Dept. of Physics, University of Wuppertal, D-42119 Wuppertal, Germany}
\affiliation{Deutsches Elektronen-Synchrotron DESY, Platanenallee 6, D-15738 Zeuthen, Germany}

\author{R. Abbasi}
\affiliation{Department of Physics, Loyola University Chicago, Chicago, IL 60660, USA}
\author{M. Ackermann}
\affiliation{Deutsches Elektronen-Synchrotron DESY, Platanenallee 6, D-15738 Zeuthen, Germany}
\author{J. Adams}
\affiliation{Dept. of Physics and Astronomy, University of Canterbury, Private Bag 4800, Christchurch, New Zealand}
\author{S. K. Agarwalla}
\thanks{also at Institute of Physics, Sachivalaya Marg, Sainik School Post, Bhubaneswar 751005, India}
\affiliation{Dept. of Physics and Wisconsin IceCube Particle Astrophysics Center, University of Wisconsin{\textemdash}Madison, Madison, WI 53706, USA}
\author{J. A. Aguilar}
\affiliation{Universit{\'e} Libre de Bruxelles, Science Faculty CP230, B-1050 Brussels, Belgium}
\author{M. Ahlers}
\affiliation{Niels Bohr Institute, University of Copenhagen, DK-2100 Copenhagen, Denmark}
\author{J.M. Alameddine}
\affiliation{Dept. of Physics, TU Dortmund University, D-44221 Dortmund, Germany}
\author{N. M. Amin}
\affiliation{Bartol Research Institute and Dept. of Physics and Astronomy, University of Delaware, Newark, DE 19716, USA}
\author{K. Andeen}
\affiliation{Department of Physics, Marquette University, Milwaukee, WI 53201, USA}
\author{G. Anton}
\affiliation{Erlangen Centre for Astroparticle Physics, Friedrich-Alexander-Universit{\"a}t Erlangen-N{\"u}rnberg, D-91058 Erlangen, Germany}
% ...
